\documentclass[%
 reprint,
nofootinbib,
 amsmath,amssymb,
 aps
]{revtex4-1}
\pdfoutput=1
\usepackage{graphicx}
\usepackage[utf8]{inputenc}
\usepackage{dcolumn}
\usepackage{bm}

\usepackage{verbatim}
\usepackage{color,ulem}

\usepackage[utf8]{inputenc}

\def\be{\begin{equation}}
\def\ee{\end{equation}}
\def\bea{\begin{eqnarray}}
\def\eea{\end{eqnarray}}

\begin{document}

\preprint{}[IFT-UAM/CSIC-19-24]

\title{\huge 
Linear Viscoelastic Dynamics in Holography
}

\author{Tomas Andrade}
 \email{tandrade@icc.ub.edu}
\affiliation{Departament de Fisica Quantica i Astrofısica, Institut de Ciencies del Cosmos,
Universitat de Barcelona, Martı i Franques 1, E-08028 Barcelona, Spain}

\author{Matteo Baggioli}%
 \email{mbaggioli@physics.uoc.gr}
 \thanks{\url{www.thegrumpyscientist.com} }
\affiliation{Instituto de Fisica Teorica UAM/CSIC,
c/ Nicolas Cabrera 13-15, Cantoblanco, 28049 Madrid, Spain
}%

 \author{Oriol Pujol{\`a}s}%
 \email{pujolas@ifae.es}
\affiliation{Institut de F\'isica d'Altes Energies (IFAE), The Barcelona Institute of Science and Technology (BIST)\\
Campus UAB, 08193 Bellaterra, Barcelona.
}%

\begin{abstract}

We study the mechanical response under time-dependent sources of a simple class of holographic models that exhibit viscoelastic features.
The ratio of viscosity over elastic modulus defines an intrinsic relaxation time scale – the so-called Maxwell relaxation time $\tau_M$, which has been identified traditionally with the relaxation time scale.  We compute explicitly the relaxation time in our examples and find that it differs from $\tau_M$.
At high temperatures $\tau_M$ over-estimates the actual relaxation time, although not by much and moreover it still captures reasonably well the temperature behaviour. At sufficiently low temperatures the situation is reversed: $\tau_M$ underestimates the actual relaxation time, in some cases quite drastically. Moreover, when $\tau_M$ under-estimates the real-time response exhibits an overshoot phenomenon before relaxation. We comment on the $T = 0$ limit, where the relaxation is power-law because our models exhibit criticality.

\end{abstract}

\pacs{Valid PACS appear here}
\maketitle

\section{Introduction}  
The difference between a rigid solid and a flowing liquid appears to be very neat and clear in our everyday experience. Despite the apparent simplicity of the question, classifying objects into solids or liquids is less trivial than expected and sometimes it can take considerable time \cite{Edgeworth_1984}. All natural materials exhibit both viscous and elastic characteristics when undergoing deformation. The interplay of elasticity and viscous behaviour takes the name of \textit{viscoelasticity} and it represents an important topic not only from the theoretical point of view but especially for applications and technological developments \cite{lakes1998viscoelastic,christensen2003theory,gutierrez2013engineering}.
A description of viscoelasticity from fundamental principles is far from being conclusive. Part of the reason for this is that the hydrodynamic framework is not able to capture the elastic response in a controlled way. On the other hand, there are effective field theory methods which allow to describe elasticity, however dissipation and finite temperature corrections can't be incorporated in a straightforward way neither. As a matter of fact most of the understanding relies on very empirical and phenomenological models such as the Maxwell model, 
the Kelvin–Voigt model and extensions thereof (see below).\\

Holography has recently been established as a valuable tool for hydrodynamics, especially for strongly correlated or strongly coupled fluids \cite{Policastro:2002se,Son:2007vk,Baier:2007ix,Hartnoll:2016apf}. More recently, progress  has been made to introduce elastic effects due to the spontaneous breaking of translational invariance and coexisting with the viscous properties implied by the presence of a black hole horizon, \textit{i.e.} finite temperature. In this direction, a promising direction to describe viscoelastic materials within holography is given by `massive gravity' theories \cite{Baggioli:2014roa,Alberte:2015isw}\footnote{See \cite{Amoretti:2017frz,Baggioli:2018bfa,Andrade:2017cnc,Li:2018vrz,Grozdanov:2018ewh} for other possibilities.}. These models break translational invariance while retaining homogeneity and thus allow to deepen the analysis. It has already been shown \cite{Alberte:2015isw,Alberte:2016xja,Alberte:2017cch,Alberte:2017oqx} that such theories possess both a finite elastic shear modulus $\mu$ aside from a finite shear viscosity coefficient $\eta$. Additionally, these models contain  propagating transverse phonons with a speed of sound that complies with standard elasticity theory \cite{Alberte:2017cch,Alberte:2017oqx}.\\
\begin{figure}[hbtp]
\centering
\includegraphics[width=0.8\linewidth]{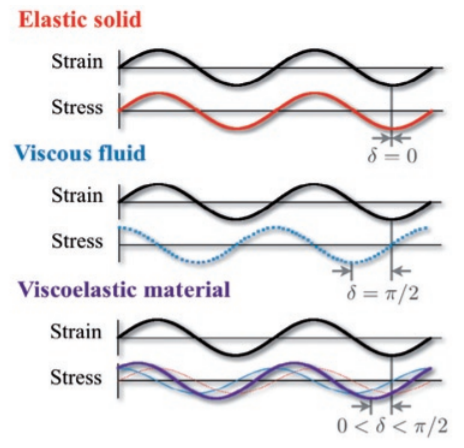}

\vspace{0.4cm}

\includegraphics[width=0.8\linewidth]{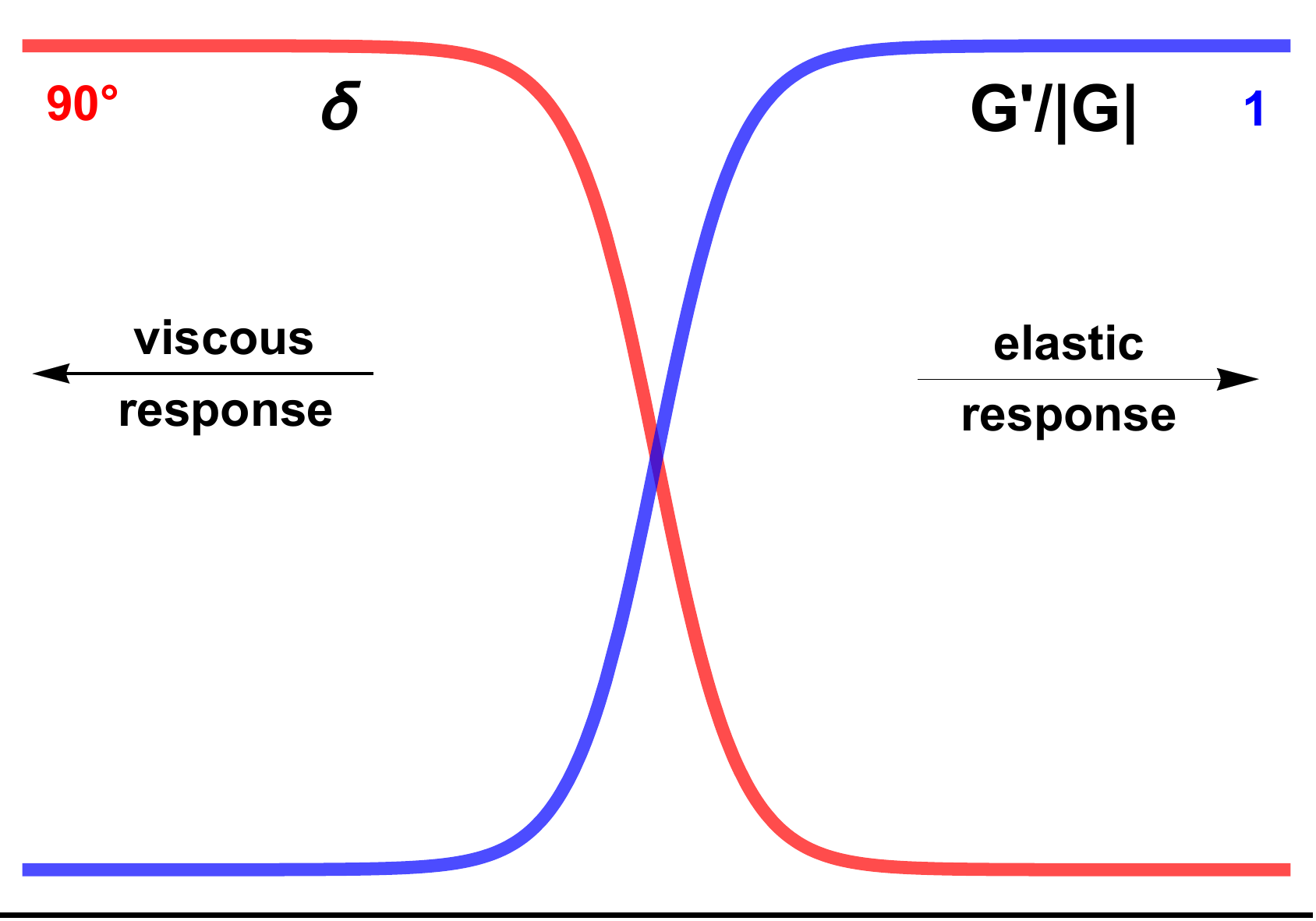}
\caption{Typical viscoelastic response. An oscillatory strain is applied to the system and an oscillatory shear stress is measured. $\delta$ is the phase between the source and the response. $\delta=90$ degrees is a purely viscous response while $\delta=0$ is a purely elastic response. Viscoelastic materials lie in the middle. A similar classification criterium is the ratio between the real part of the complex dynamical modulus and its absolute value $G'/|G|$. For $G'/|G|=0$ the response is purely viscous while for $G'/|G|=1$ it is purely elastic. %
The left panel is taken and adapted from \cite{reffig}.}
\label{sketchfig}
\end{figure}\\
Beyond the definition and the direct measurement of the viscosity and the elastic modulus, viscoelasticity gives very peculiar imprints on the real time (or finite frequency) response of a system under deformation. A typical experimental test is the so-called \textit{dynamic mechanical analysis} which consists in applying to the sample a small oscillatory stress $\sigma$ and measuring the resulting strain $\epsilon$. More precisely considering a dynamical load of the form $\sigma(t)=\sigma_0 \cos (\omega t)$ applied to a viscoelastic sample, the response will be a strain of the form $\epsilon(t)=\epsilon_0 \cos (\omega t-\delta)$ where $\delta$ is called the \textit{loss angle} of the material. Expanding the above expression we obtain:
\begin{equation}
\epsilon(t)\,=\,\underbrace{\epsilon_0\,\cos (\omega t) \cos \delta}_{elastic}\,+\,\underbrace{\epsilon_0\,\sin (\omega t) \sin \delta}_{viscous}
\end{equation}
where the first term is completely in phase with the input while the second term is out of phase. The first term is the elastic non dissipative response; setting $\delta=0$ indeed the stress and the strain are completely in phase. On the contrary the second term is the viscous response and setting $\delta=90$ degrees the stress and the strain are completely out of phase. In order to motivate such statement it is easy to calculate the work done in stressing a material (per unit volume) which is given by $W=\int \sigma d\epsilon$. A short computation shows that the energy loss is:
\begin{equation}
W\,=\,\pi\,\sigma_0\,\epsilon_0\,\sin \delta
\end{equation}
which indeed reflects the fact that $\delta$ measures a dissipative contribution, \textit{i.e.}  for $\delta=0$  no energy is dissipated and the process is  reversible -- the response is  purely elastic.\\
More generically we can say that the elastic response is proportional to the strain itself $\epsilon(t)$ while the viscous response is proportional only to the strain rate $\dot{\epsilon}(t)$. It is indeed standard to assume that a liquid does not react to a static strain deformation.\\
Sometimes it is convenient to regard the strain as the input and the stress as the output\footnote{Notice this is also doable in experimental setups despite it is less under control than fixing the strain and reading the induced stress.}. Given this assumption, we can re-write the induced stress as:
\begin{equation}
\sigma(t)\,=\,\epsilon_0\,\left[G'\,\cos(\omega t)\,+\,G'' \sin (\omega t)\,\right]\,=\,G'\,\epsilon(t)\,+\,\frac{G''}{\omega}\,\dot{\epsilon}(t)
\end{equation}
where we defined:
\begin{equation}
	 G'\,=\,\frac{\sigma_0}{\epsilon_0}\,\cos \delta \,,\quad G''\,=\,\frac{\sigma_0}{\epsilon_0}\,\sin \delta.
\end{equation}
It is then customary to represent the linear relation between the oscillating stress and strain in Fourier space defining the \textit{complex dynamic modulus} $G$ as
\begin{equation}
\label{G def}
\sigma(\omega)\,=\,G(\omega)\,\epsilon(\omega),\quad\quad {\rm with} \quad\quad G=G'\,+\,i\,G''
\end{equation}
The real part $G'$ takes the name of \textit{storage modulus} and the imaginary and dissipative one $G''$ of \textit{loss modulus}%
\footnote{Note that we use the standard notation in condensed matter physics, 
where prime and double prime on $G$ denote
its real and imaginary parts, not derivatives with respect to the argument. 
Let us also mention that often in the literature, the study of how matter 'flows' or responds to time dependent strains is referred to as {\it rheology}.}. 
In purely elastic materials the response is in phase and therefore 'instantaneous'. In purely viscous fluids, on the contrary, the response has a phase shift of $\delta=90$ degrees which corresponds to $G'=0$ (since $\tan \delta= G'/G''$); viscoelastic materials lie in between. At the same time we can classify the response using the dimensionless ratio $G'/|G| = \cos\delta$ between the storage modulus and the absolute value of the complex dynamic modulus. A purely viscous material would have such ratio equal to zero, while in a purely elastic one that number will be maximum, \textit{i.e.} $G'/|G|=1$.

%
Given the expression \eqref{G def} it appears clear that the complex moduli $G(\omega)$ can be identified as the (retarded) Green function for the shear stress operator $\sigma \equiv T_{xy}$. Therefore the following identification holds
\begin{equation}
\label{G and G}
	G(\omega) \,\equiv\, \mathcal{G}^{\textrm{(R)}}_{T_{xy}T_{xy}}(\omega)
\end{equation}
and it will be very useful for extracting the complex moduli within the holographic setup. Indeed from the operational point of view the quantity \eqref{G and G} is what we will compute.
That said, we can easily identify from the complex modulus the (static) shear elastic modulus $\mu$ and the 
shear viscosity $\eta$ as:
\begin{equation}
\mu\,\equiv\, G'(0)\,,\quad \eta\,\equiv\,-\,\lim_{\omega \rightarrow 0}\,\left(\frac{G''(\omega)}{\omega}\right)
\end{equation}
which together determine the low frequency behaviour of the complex modulus itself:
\begin{equation}
\label{hydro}
G(\omega)\,=\,\mu\,-\,i\,\omega\,\eta\,+\,\mathcal{O}\left(\omega^2\right)~.
\end{equation} 
It is illustrative to discuss one of the simplest phenomenological models, the so-called Maxwell viscoelastic fluid model.
The response under homogeneous but time dependent source  is modelled by assuming that the relation between instantaneous strain and stress follow the linear differential equation
\begin{equation} \label{Maxwell}
	\sigma(t) + \frac{\eta}{G_\infty} \dot \sigma(t) = \eta\; \dot \epsilon(t) \quad\qquad {\rm(Maxwell~model)}
\end{equation}
with $\eta$ the viscosity and  ${G_\infty} $ a parameter that is immediately recognized as the effective elastic shear modulus at large frequencies, sometimes also referred to as the instantaneous elastic shear modulus. The model then interpolates between fluid behaviour (zero response to a static strain) and solid like behaviour at large frequencies. The fact that liquids behave like solids at frequencies higher than the so-called Frenkel frequency $\omega_F \gg 1/ \tau_M = G_\infty/\eta$ is known since long time \cite{doi:10.1021/j150454a025} and it can be easily experienced by diving from high heights.

It is worth showing the complex shear modulus in frequency space,
\begin{equation}\label{Maxwellw}
	G(\omega) = \frac{\eta \,\omega}{i + \eta/G_\infty \, \omega }
\end{equation}
which clearly exhibits $G_\infty = \lim_{\omega \to \infty} G(\omega)$. Given that this is nothing but a Green function, it is clear that this signals the presence of a quasi-resonance -- we shall stick to the naming {\it quasi normal model} (QNM) that is more common  in the AdS/CFT literature. 
The Maxwell model contains one single QNM, located at
\begin{equation}
\label{omega M}
	\omega_M = - i\, \frac{G_\infty}{\eta}~.
\end{equation}
Hence this model exhibits  relaxation to the steady state with typical time dependence $\sim e^{- t / \tau_M}$ with the Maxwell relaxation time
\begin{equation}
\label{tauM}
	\tau_M = \frac{\eta}{G_\infty}~.   \;\;\;\quad {\rm(Maxwell~relaxation~time)}
\end{equation}

From our perspective, there are two important lessons to extract from this model. First, that materials (fluid or not) already possess a relaxation time scale as given by \eqref{tauM} and which is important for dynamical viscoelastic response experiments. Let us add that for solids, there is a similar quantity that can be obtained, out of the static elastic modulus $\mu$.  Second, the QNMs in the $ T_{xy}T_{xy} $ correlator (poles in the complex modulus) are also naturally compared to the inverse Maxwell relaxation  time $1/\tau_M$. In generic materials, one can expect i) that there are several QNMs and ii) that their locations {\it do not} exactly coincide with $-i/\tau_M$, not even the lightest one (the one with  smallest imaginary part). 
In the Maxwell model \eqref{Maxwell} the location of the QNM coincides with the inverse Maxwell time, $\tau_M^{-1}=G_\infty/\eta$,
however this is due to the model simplicity. It is easy to break the coincidence by extending the model (as we show in Sec \ref{sec:toy} this can be done while having still a single QNM).
In the holographic models below we will also find that the relaxation rate from the lightest QNM differs from 
$\tau_M^{-1}$, and this has an interesting impact in the viscoelastic response. \\


Once $G(\omega)$ is known we can also compute the real-time response of a system under an external and time dependent source. 
Viscoelastic materials exhibit several interesting features in the real-time response to a time dependent deformation. Among these, we shall mainly focus on :
\begin{itemize}
\item \textit{Stress Relaxation}.\\ 
It is the decrease in stress in response to a constant strain generated in the structure, which can be seen as a delay of the material in reaching the steady-state. Its time dependence can typically be power-law or exponential, and in the latter case one can identify a relaxation time $\tau$. In the  phenomenological Maxwell model the relaxation time is simply the ratio between the elastic modulus and the viscosity $\tau_M=\eta/G_{\infty}$ where the latter is the value of $G'(\omega)$ at infinite frequency.
\item \textit{Stress Overshoot}.\\
It refers to the appearance of a transient excess stress before reaching the (non-zero) steady state. 
It is typically seen in start-up experiments, (e.g. when the source is turned on  at a given time) 
and has been reported experimentally in a range of materials, from amorphous polymers and metallic glasses to foams and gels (see \cite{falk2011deformation} for a review).
More precisely the \textit{stress overshoot} is characterized by the following temporal sequence: 
the stress $\sigma$ first increases linearly with time, until it reaches a maximum value and then decreases
towards it steady-state constant value. Despite several explanatory attempts a complete understanding is still missing, see 
\cite{divoux2011stress}.
\end{itemize}

In the remainder of this paper we shall study the viscoelastic properties of the massive gravity models of \cite{Baggioli:2014roa,Alberte:2017oqx} by is computing the complex modulus and the response under time dependent sources. This opens the possibility of studying viscoelastic effects in strongly coupled materials or systems at quantum criticality. In the following we will show that the holographic setup at hand exhibits standard viscoelastic properties, starting by a clear \textit{stress relaxation}. As expected, the relaxation time is controlled at finite temperature by the least damped QNM, specifically by the inverse of its imaginary part, 
$$
\tau_{rel}\,=\,-\, 1/ Im\left(\omega_{QNM}^*\right)~.
$$
Moreover, in general this relaxation time does not  coincide with the Maxwell relaxation time $\tau_M=\eta/G_{\infty}$ (even though in some cases they are within the same order of magnitude, and even the temperature dependence is similar). Additionally, we find that the stress overshoot phenomenon appears when the relaxation time is larger than the Maxwell time, $\tau_{rel}>\tau_M$. 
Finally, at $T=0$ the stress relaxation becomes power-law and governed by the formation of a branchcut on the imaginary axis. 
The power is simply dictated by the conformal dimension of the stress operator on the near-horizon IR zero temperature geometry which is derived in details in Appendix \ref{appIR}.\\

This paper is organized as follows: in section \ref{sec:setup} we will set the stage presenting the holographic models we will consider, section \ref{sec:reho} contains the main results of the manuscript, finally section \ref{sec: conclusions} summarize our findings and propose some future directions.

\section{Holographic setup}\label{sec:setup}

We consider generic \textit{viscoelastic} holographic massive gravity models \cite{Baggioli:2014roa,Alberte:2015isw}: 
\begin{equation}\label{action}
S\,=\, M_P^2\int d^4x \sqrt{-g}
\left[\frac{R}2+\frac{3}{\ell^2}- \, m^2 V(X,Z)\right]
\end{equation}
which are inspired from the original model \cite{Andrade:2013gsa}. We define $\mathcal{I}^{IJ}=g^{\mu\nu} \,\partial_\mu \phi^I \partial_\nu \phi^J$ and $X=Tr[\mathcal{I}],Z=det[\mathcal{I]}$. The dual field theory has been shown to possess viscoelastic features and was studied in detail in \cite{Baggioli:2015zoa,Baggioli:2015gsa,Baggioli:2015dwa,Alberte:2016xja,Alberte:2017cch,Alberte:2017oqx}. We consider 4D asymptotically AdS black hole geometries in Eddington-Filkenstein (EF) coordinates:
\begin{equation}
\label{backg}
ds^2=\frac{1}{u^2} \left[-f(u)\,dt^2-2\,dt\,du + dx^2+dy^2\right]
\end{equation}
where $u\in [0,u_h]$ is the radial holographic direction spanning from the boundary to the horizon, defined through $f(u_h)=0$. The $\phi^I$ are the St\"uckelberg scalars which admit  a radially constant profile $\phi^I=x^I$ with $I=x,y$ and the emblackening factor takes the simple form:
\begin{equation}\label{backf}
f(u)= u^3 \int_u^{u_h} dv\;\left[ \frac{3}{v^4} -\frac{m^2}{v^4}\, 
V(v^2,v^4) \right] \, ,
\end{equation}
The corresponding temperature of the dual QFT reads:
\begin{equation}
T=-\frac{f'(u_h)}{4\pi}=\frac{6 -  2 m^2 V\left(u_h^2,u_h^4 \right) }{8 \pi u_h}~.
\end{equation}
while the entropy density is simply $s=2\pi/u_h^2$. The heat capacity can be simply obtained as $c_v=T ds/dT$ and was studied in \cite{Baggioli:2015gsa,Baggioli:2018vfc}.\\

The deformation and flow properties of the model are encoded in the $T_{xy}$ operator which is dual, by the holographic dictionary, to the bulk deformation $\delta g_{xy}\equiv h_{xy}$.
At zero momentum $k=0$ the linearized equation for $h_{xy}$ decouples and in EF coordinates reads:
\begin{equation}
h_{xy} \left(-\frac{2 m^2 V_X}{f}-\frac{2 i \omega}{u f}\right)+h_{xy}' \left(\frac{f'}{f}+\frac{2 i \omega}{f}-\frac{2}{u}\right)+h_{xy}''=0\label{EQ}
\end{equation}
where $V_X\equiv \partial_X V(X,Z)$ and primes denote radial derivatives. 
The UV asymptotic behaviour of the $h_{xy}$ field is:
\begin{equation}
h_{xy}\,=h_{xy\,(l)}(\omega)\,(1\,+\,\dots)\,+\,h_{xy\,(s)}(\omega)\,u^{3}\,(1\,+\,\dots)
\end{equation}
The AdS/CFT dictionary allows us to express the Green's function of the stress tensor as
\begin{equation}
 \mathcal{G}^{\textrm{(R)}}_{T_{xy}T_{xy}}(\omega)\,=\,\frac{2\,\Delta-d}{2}\,\frac{h_{xy\,(s)}(\omega)}{h_{xy\,(l)}(\omega)}\,=\,\frac{3}{2}\frac{h_{xy\,(s)} (\omega) }{h_{xy\,(l)} (\omega)}\label{green}
\end{equation}
%
In order to obtain the retarded correlator, one needs to impose ingoing boundary conditions at the horizon $u=u_h$, which 
in EF coordinates corresponds to regularity of $h_{xy}$. We also compute the spectrum of quasi-normal modes 
by solving the boundary value problem resulting from setting setting the source to zero. They correspond
to the poles of the Green's function of the dual theory. 

In the following we will consider potentials of the form $V(X,Z)=X^n$, $n>5/2$ and $V(X,Z) = Z^m$, $m>5/4$
which realize the spontaneous breaking of translational invariance, and exhibit both elastic and viscous 
response \cite{Alberte:2017oqx}.  For later use, we recall the expressions for the energy density in these 
models, this being 

\begin{equation}
\epsilon\,=\,\frac{1}{u_h^3}\,+\,m^2\,\int_0^{u_h} \frac{V(u^2,u^4)}{u^4}\,du
\end{equation}
%
%
%
%

We will refer to the previous class of models as \cite{Alberte:2015isw,Alberte:2016xja}:
\begin{itemize}
\item Viscoelastic solid model: $V(X,Z)=X^n$, $n>5/2$ 
\item Viscoelastic fluid model: $V(X,Z) = Z^m$, $m>5/4$
\end{itemize}
We take as an operational distinction between a fluid and a solid the presence of a finite static shear modulus $\mu \neq 0$. 
In other words, we will call solids all those configurations which posses a finite response to a zero frequency 
mechanical deformation. The adjective ''viscoelastic'' emphasizes that in both models there is a 
competition between $G'(\omega)$ and $G''(\omega)$ at finite $\omega$ due to the 
non-trivial frequency dependence of the response. 

\section{Holographic Rheology}
\label{sec:reho}

Rheology is the study of the deformation and flow of materials, both solids, liquids and especially viscoelastic ones. 
It combines the analysis of the viscous and elastic properties of a given material and it is based, at least at linear level, 
on the relation between the stress $\sigma(\omega)$ in a material and its consequent mechanical deformation, 
\textit{i.e.} the strain $\epsilon(\omega)$. In a more formal way it is based on the relation \eqref{G def}
involving the \textit{complex dynamic modulus} $G(\omega)$. This can be related to the correlator of particular 
spatial components of the stress tensor as in \eqref{G and G}, which facilitates our holographic treatment. 
In our language the dynamical modulus is simply encoded in the Green function of the shear operator $T_{xy}$ which 
can be extracted numerically from \eqref{green} using the standard holographic techniques.

Once $G(\omega)$ is obtained we can compute the real-time response of a system under an external and time dependent source $\sigma(t)$. To this end, we will Fourier transform back to the time domain and write
\begin{equation}
\label{convolution}
\sigma(t)\,=\,Re\left[\int_{-\infty}^\infty\,e^{-\,i\,\omega\,t}\,G(\omega)\,\tilde{\epsilon}(\omega)\right]
\end{equation}
where $\tilde{\epsilon}(\omega)$ is simply the Fourier transform of the external strain source $\epsilon(t)$. 
For simplicity we will consider the following external sources:
\begin{itemize}
\item \textit{Logistic function}.
\begin{equation}
\mathfrak{L}(t,\beta)\,\equiv\,\frac{1}{1\,+\,e^{2\,\beta\,t}}
\end{equation}
It represents a smooth deformation of a step function and its Fourier transform is:
\begin{equation}
\tilde{\mathfrak{L}}(\omega,\beta)\,=\,\sqrt{\frac{\pi }{2}} \,\delta (\omega )+\frac{i \sqrt{\frac{\pi }{2}}\, \text{csch}\,\left(\frac{\pi  \omega }{2 \beta }\right)}{2 \,\beta }
\end{equation}

\item \textit{Ramp function}.
\begin{equation}
\label{ramp source}
\mathfrak{R}(t)\,\equiv\,\begin{cases}
t\quad \textit{for}\quad t>0\\
0\quad \textit{otherwise}
\end{cases}
\end{equation}
whose Fourier transform is:
\begin{equation}
\tilde{\mathfrak{R}}(\omega)\,=\,-\frac{1}{\sqrt{2 \,\pi }\, \omega ^2}-i\, \sqrt{\frac{\pi }{2}}\, \delta '(\omega )
\end{equation}
\end{itemize}

\subsection{Fluids}\label{sec:Z}

We start by considering the viscoelastic fluid model $V(X,Z)=Z^n$.
For concreteness we restrict ourselves to the  potential
\begin{equation}
\label{V_Z2}
V(X,Z) =  Z^2
\end{equation}
which corresponds to consider a dual QFT in a fluid phase \cite{Alberte:2015isw} which exhibits spontaneous 
breaking of translational invariance. Other potentials of the form 
$V(X,Z) = \mathcal{V}(Z)=Z^n$, with $n>5/4$ give similar results\footnote{Nevertheless notice that this setup can provide also a mechanism for the explicit breaking of translations (for example choosing $\mathcal{V}(Z)=Z$) and therefore a finite DC conductivity still having zero elastic modulus $\mu=0$ \cite{Alberte:2015isw}.}. 
In particular, since $V_X=0$, it is easy to prove \cite{Alberte:2016xja} that for this model:
\begin{equation}
\mu\,=\,G'(0)\,=\,0\,,\quad \frac{\eta}{s}\,=\,\frac{1}{4\,\pi}
\end{equation}
which is just a consequence of the fact that the shear component of the graviton remains massless. 
In other words, this model has no static elastic response but purely a viscous one, 
which justifies the term ``fluid".
\begin{figure}[th]
\centering
\includegraphics[width=0.8 \linewidth]{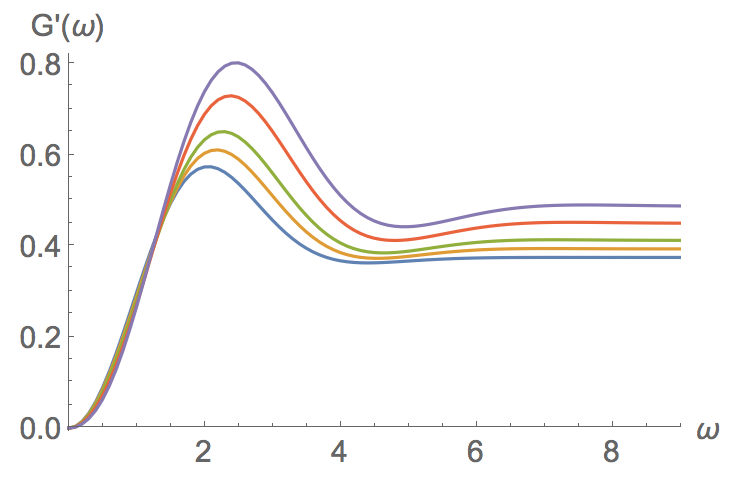}

\vspace{0.4cm}

\includegraphics[width=0.8\linewidth]{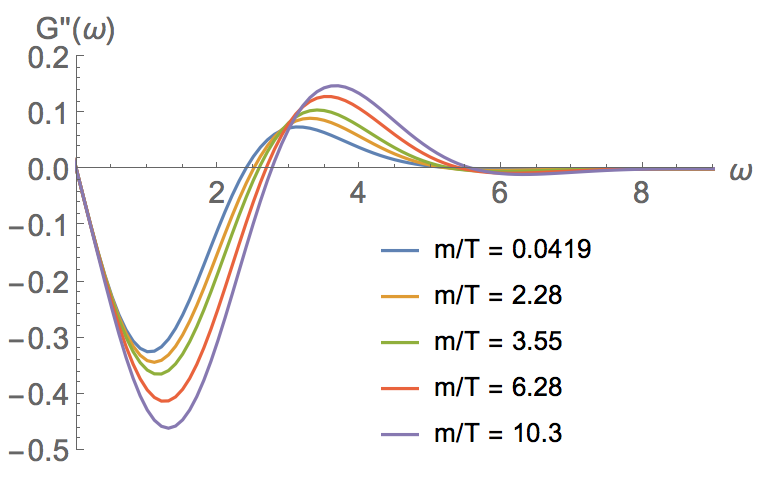}
\caption{Green's function as a function of frequency for $V = Z^2$ for 
various values of $m/T$.}%
\label{fig:G_Z2}
\end{figure}\\
We start by discussing the finite frequency response of the complex dynamical modulus $G(\omega)$. 
The results are shown in fig.\ref{fig:G_Z2}. Note that by time reflection symmetry, $G'(-\omega) = G'(\omega)$, 
$G''(-\omega) = -G''(\omega)$, so we only show our results for $\omega \geq 0$. The same argument applies to the structure 
of QNMs which appear in pairs symmetric under ${\rm Im} \omega_{QNM} \to - {\rm Im} 
\omega_{QNM}$. 
As already described, at low frequencies we find $\mu = G'(0)=0$ while the viscosity saturates 
the KSS bound $\eta/s=1/4\pi$ \cite{Kovtun:2004de}. 
The imaginary part $G''(\omega)$ goes to zero at large frequency, so no dissipation is present in this regime. 
Physically, this means that if the system is perturbed at a time scale much shorter than its typical 
relaxation time, it is not able to dissipate.
It is interesting to note that at large frequencies, the real part of $G(\omega)$ takes a constant value which 
we denote $G_{\infty} \equiv G'(\omega=\infty)$. As we show in appendix \ref{app:ginf}, this can be computed 
analytically in terms of the energy density as:
\begin{equation}
\label{Ginf E}
G_{\infty}\,= \frac{3}{8} \epsilon
\end{equation}
\noindent independent of the specific form of the potential.

\begin{figure}[th]
\centering
\includegraphics[width=0.8 \linewidth]{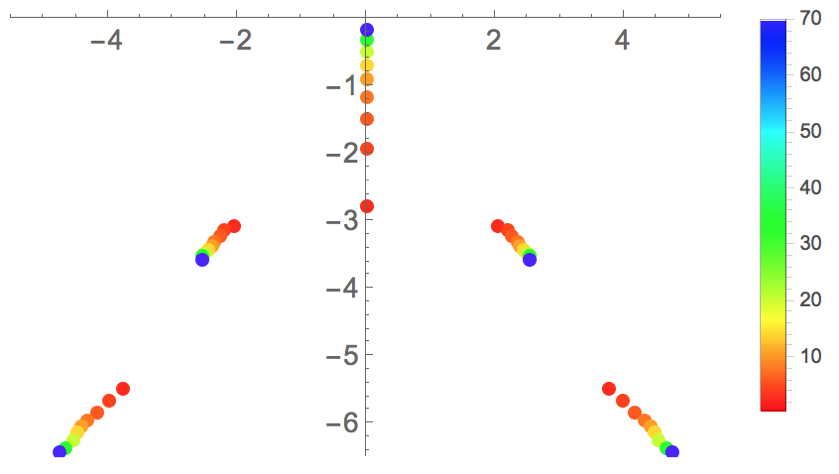}

\vspace{0.4cm}

\includegraphics[width=0.8 \linewidth]{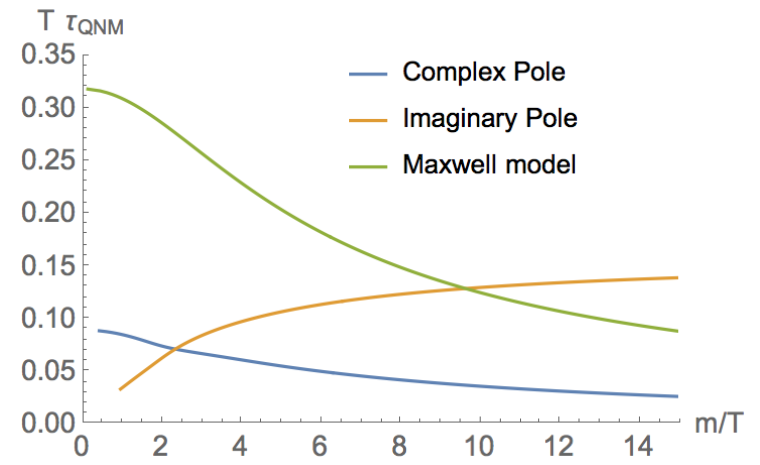}

\caption{{\bf Top:} QNMs in the complex plane for $V=Z^2$ and varying $m/T$. {\bf Bottom:} Relaxation times associated to the lightest complex and purely imaginary QNMs as a function of $m/T$ (the actual relaxation time of the system is the largest of the two). For comparison, we show also the Maxwell relaxation time defined as $\tau_M\equiv \eta/G_\infty$. At  $m/T\gtrsim 10$, the Maxwell relaxation time becomes smaller than the QNM relaxation time scale.}%
\label{fig:QNM_Z2}
\end{figure}

In the intermediate frequency range we observe a non-trivial frequency dependence and, in particular, a 
clear resonance in $G'(\omega)$. Not surprisingly such a resonance is also visible in the QNMs spectrum of the 
system which is shown for different temperatures in fig.\ref{fig:QNM_Z2}. Note in particular that the position 
of the isolated complex pole matches quite well the position of the peak in the real part of the complex modulus.\\

From the positions of the QNMs, and in particular from the imaginary part of the least damped one, we can read off
the relaxation time which is shown in the right panel of fig.\ref{fig:QNM_Z2}. Interestingly, at a certain value $m/T \approx  2.5$ 
there is a crossover between the two dominating poles. 
Upon further lowering the temperature, a series of purely imaginary poles accumulate on the imaginary 
axes and form a branch cut at exactly zero temperature as usual in models with an extremal horizon at $T=0$, see \cite{Faulkner:2009wj}. 

\begin{figure}[th]
\centering
\includegraphics[width=0.8 \linewidth]{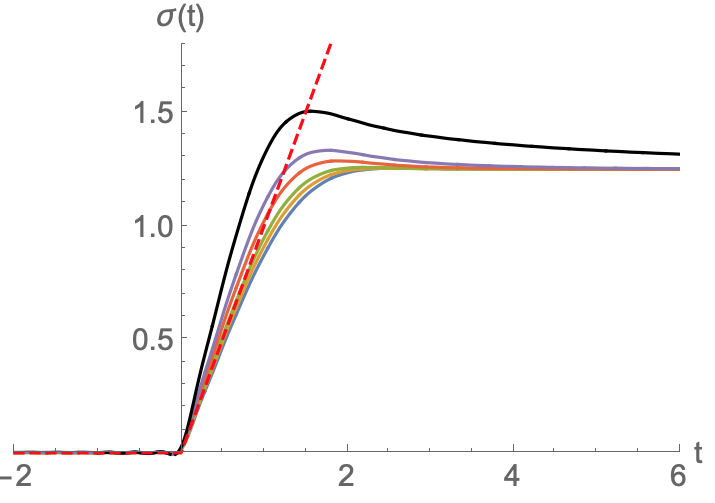}

\vspace{0.4cm}

\includegraphics[width=0.8 \linewidth]{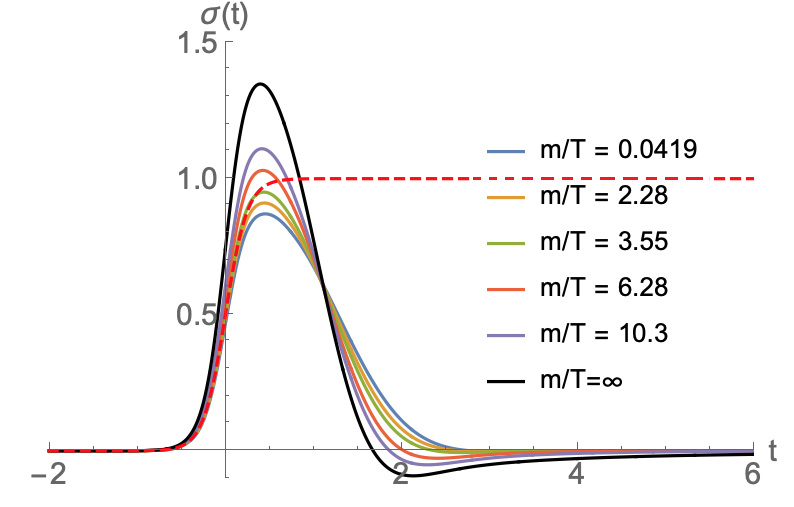}
\caption{Real time responses for the potential $V=Z^2$, for a ramp source (top) and a logistic source (bottom). The overshoot behaviour is significant for  $m/T \gtrsim 10$. }%
\label{fig:real_time_Z2}
\end{figure}

\begin{figure}[th]
\centering
\includegraphics[width=0.8 \linewidth]{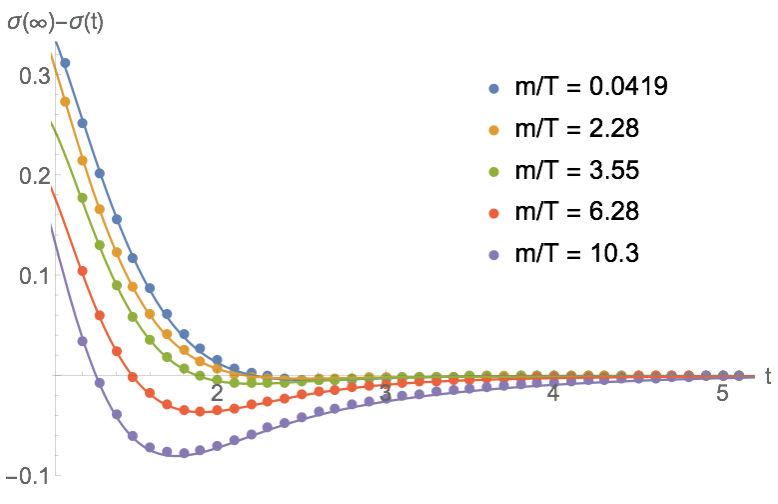}

\vspace{0.4cm}

\includegraphics[width=0.8 \linewidth]{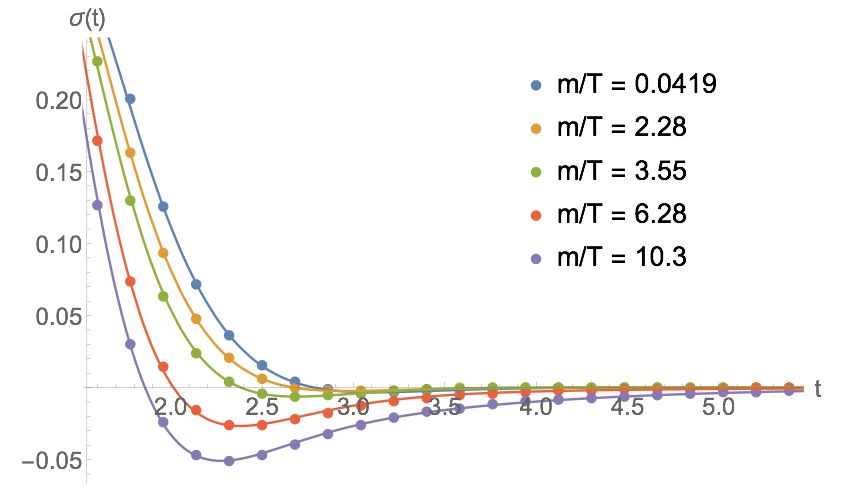}
\caption{Fit of the late time dynamics in the case of the logistic response for the $V = Z^2$ 
model to the lowest QNMs, for the case of the ramp (left) and logistic function (right). 
In the range of temperatures considered, we obtain a good fit considering both the 
purely imaginary and lightest complex modes. The data points mark our numerical data, while the solid lines correspond 
to the QNM fit. To ease visualization we have subtracted the equilibrium value.}%
\label{fig:late_Z2}
\end{figure}

Next, we consider the real time mechanical response of the system upon applying an external strain in the form 
of a ramp and step sources. We display our results in fig. \ref{fig:real_time_Z2}.
The response to the step source is simple and confirms the fluid nature of the material. In fact, we can see from the 
right panel of fig.\ref{fig:late_Z2} that the system reacts only to a gradient of the strain $\dot{\epsilon}(t)$ and only develops a 
peak close to $t = 0$. At late times where the strain is constant, no response is present, as expected for a fluid. 

The response to the ramp function is more interesting. At small $m/T$, $\sigma(t)$ follows the expected form
of early growth and (exponential) decay similar to the one observed in amorphous solids and disordered 
systems; see \cite{PhysRevB.90.140203} for an example.
As we increase $m/T$ the response departs from such a form and develops a clear peak before reaching 
the late time Newtonian plateau. Such a feature is a distinct signature of the presence of a stress overshoot, and 
is produced by the interplay between the viscous behaviour and some elastic feature present at finite frequency. 
Indeed, as shown in fig.\ref{fig:G_Z2}, despite that the shear modulus is zero, \textit{i.e.} $G'(0)=0$, the real part of the complex modulus 
exhibits a strong peak at intermediate frequency which is governed by the first resonance in the QNM spectrum and whose strength grows with
the parameter $m/T$. Correspondingly, we see that the overshoot peak grows with $m/T$. 

In both the ramp and the step setups the stress relaxes at late time towards a steady-state value (which turns out to be zero for the step source because of the absence of a finite static shear modulus $\mu$). The stress relaxation, which is exponential at finite temperature, 
is governed by the imaginary part of the least damped QNM. Indeed, we find good agreement between the real time data at late times and the QNMs, see fig.\ref{fig:late_Z2}, even at considerably small $T$, where the lightest pole is on the imaginary axis. 
In the  $T=0$ limit one expects that the tower of poles in the imaginary axis merge into a branch cut reaching down to $\omega=0$, and this turns the relaxation in real time to be power-law rather than exponential. 
While this is hard to see explicitly in the real time numerics, we find evidence of 
the formation of a branch cut on the imaginary axis\footnote{However, we do see the power-law scaling 
in our toy model to be discussed in Sec.~\ref{sec:toy}.}. 
On general grounds, this power law is controlled by the anomalous dimension that the stress tensor operator can get in the IR, which corresponds to the graviton mass in the $AdS_2$ near-horizon geometry. 
We compute this in appendix \ref{appIR}.


Finally we compare the relaxation time at finite temperature which is given by the imaginary 
part of the least damped QNM:
\begin{equation}
\tau_{rel}\,=\,-\, 1/ Im\left(\omega_{QNM}^*\right)
\end{equation}
to the Maxwell relaxation time:
\begin{equation}
\tau_M\, \equiv \,\frac{\eta}{G_{\infty}}~.
\end{equation}
Let us emphasize that this parameter is certainly present in the system simply because, just like in the Maxwell model, there is a finite $G_\infty \equiv \lim_{\omega\to\infty} G(\omega)$. In contrast with the simple Maxwell model, though, this relaxation time might not coincide with any QNM relaxation time. 
The comparison between these relaxation time scales can be seen in fig.~\ref{fig:QNM_Z2}. Quite remarkably, $\tau_M$ is rather close to the actual relaxation time. At high $T$, the difference is only a factor $\sim 3$ and the temperature dependence is the correct one. This is so until the pure imaginary QNM becomes dominant at $m/T \gtrsim 2.5$. From then on, the temperature dependence of $\tau_M$ and of the actual relaxation time $\tau_{QNM}$ differ. At $m/T\simeq10$, the Maxwell time $\tau_M$ turns from over-estimating to under-estimating the actual relaxation time.\footnote{The mismatch between the holographic relaxation time and the 
Maxwell relaxation time have been already observed in a slightly different context in 
\cite{Baggioli:2018vfc,Baggioli:2018nnp}.} 
In hindsight, the failure of $\tau_M$ is built in in the fact that at $T=0$ our model has an AdS${}_2$ critical point, meaning that the imaginary pole must become arbitrarily light or equivalently the relaxation must be power law, not exponential in time. 

The discrepancy between $\tau_M$ and $\tau_{QNM}=\tau_{rel}$ also seem to have an impact on the overshoot phenomenon seen in fig.~\ref{fig:real_time_Z2}. As we see from this figure and fig.~\ref{fig:QNM_Z2}, overshooting is present when the condition 
\be\label{overshoot}
\tau_{QNM} >  \tau_M
\ee
is met. This condition will be confirmed in the two more cases studied in the next subsections. Heuristically, it seems reasonable that if a material possesses  dynamical modes (resonances) that are light (decay slowly) then effects coming from these resonances are visible. The above condition then represents the criterion that decides whether a resonance is light or not, and it is very natural that this involves the viscoelastic moduli $\eta$, $G_\infty$.

\subsection{A simple toy model}\label{sec:toy}

In this section we construct a simple toy model for the Green function of our system. Despite its simplicity, 
the model is able to reproduce several features present in the more complex holographic setup. 
In order to mimic the $T=0$ solution we take a Green's function which contains a branch cut 
on the imaginary axis, as in the AdS$_2$ dual geometry, and a series of isolated poles. More precisely we consider:
\begin{equation}
\mathcal{G}_{toy}(\omega)\,=\,c\,+\,\sum_{k}\,\frac{b}{\omega^2\,+\,i\,\Gamma_k\,\omega\,-\,m^2_k}\,+\,a\,\left(-\,i\,\omega\right)^\alpha\,+\,\dots \label{toygreen}
\end{equation}
where we have set the residues of all the poles equal. By choosing $m_k$ and $\Gamma_k$ appropriately, one can arrange for an array of poles, such as  forming an inverse parabola-like shape like in the holographic models. 
However, it will be sufficient to consider only the least damped mode of the tower so we shall not enter into specifying this further.
\begin{figure}
\centering
\includegraphics[width=0.8\linewidth]{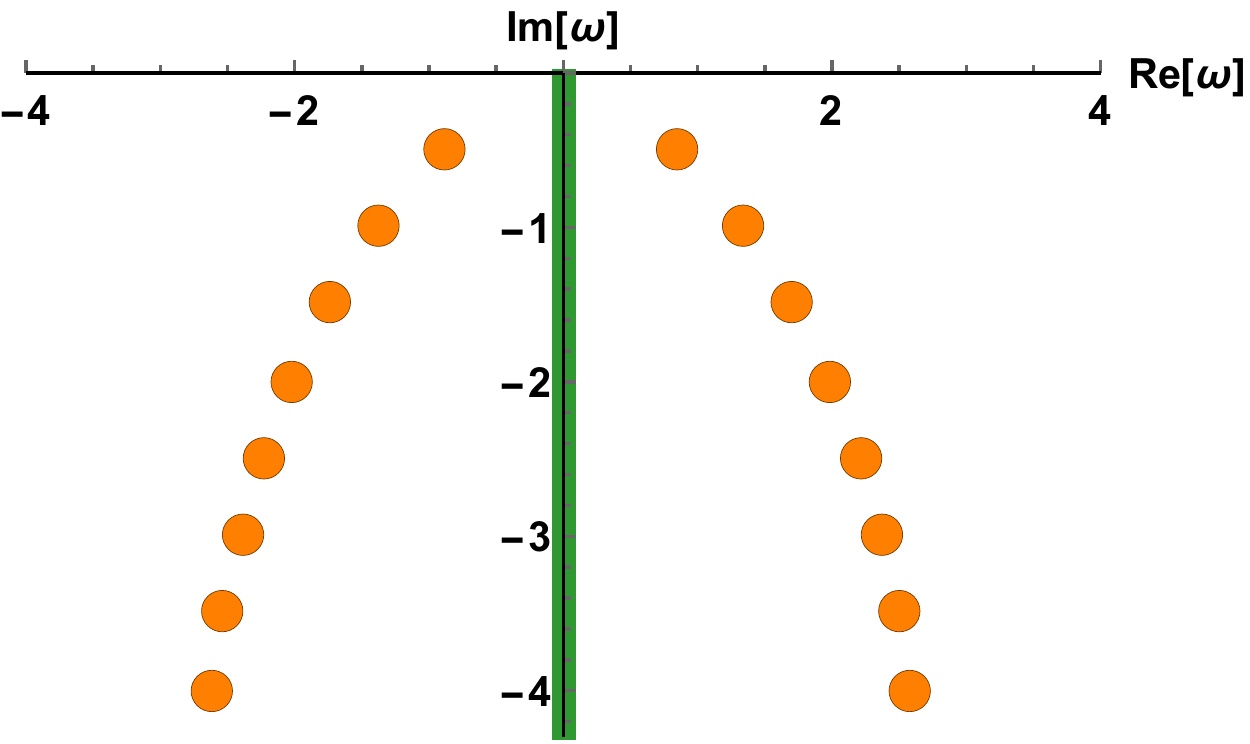}
\caption{
Analytic structure fo the toy model Green's function \eqref{toygreen}, dots indicating poles and the green line the branchcut. 
}
\label{figtoy1}
\end{figure}

\begin{figure}
\centering
\includegraphics[width=0.8\linewidth]{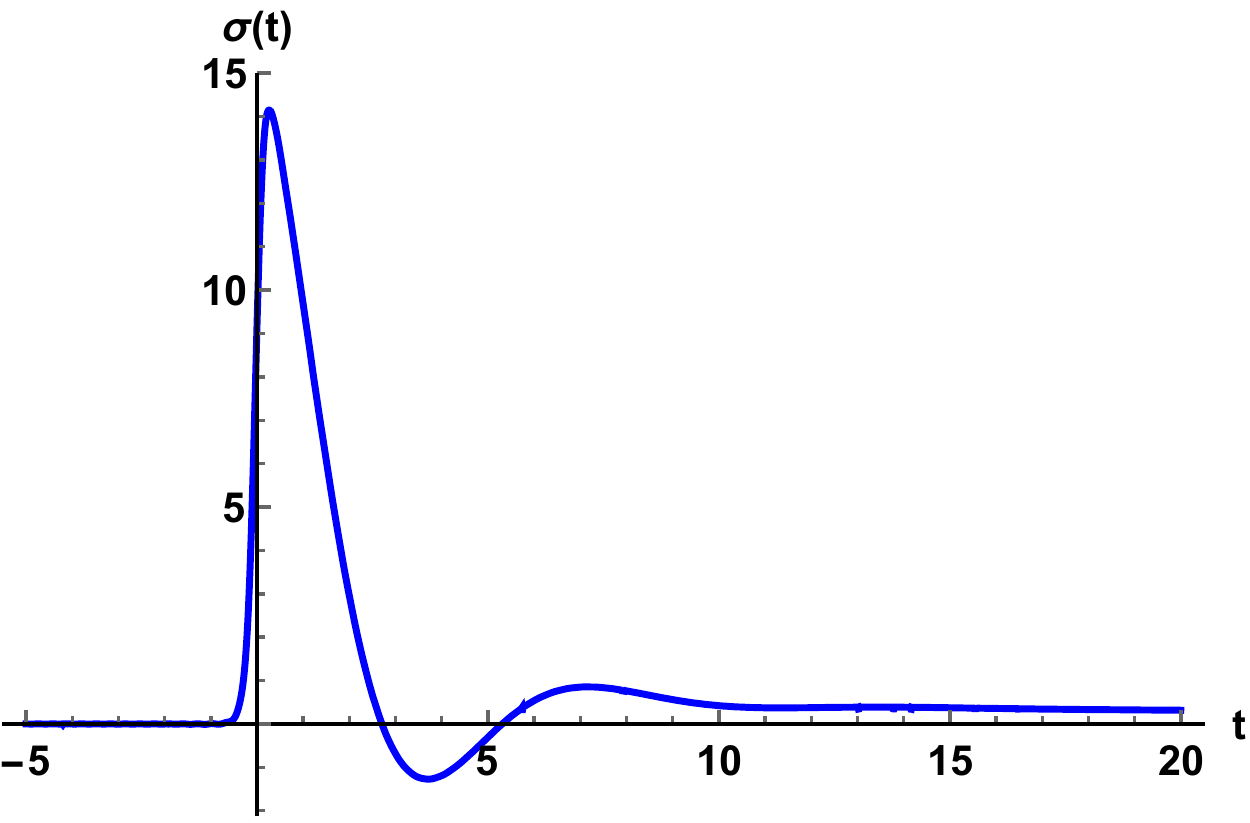}

\vspace{0.4cm}

\includegraphics[width=0.8\linewidth]{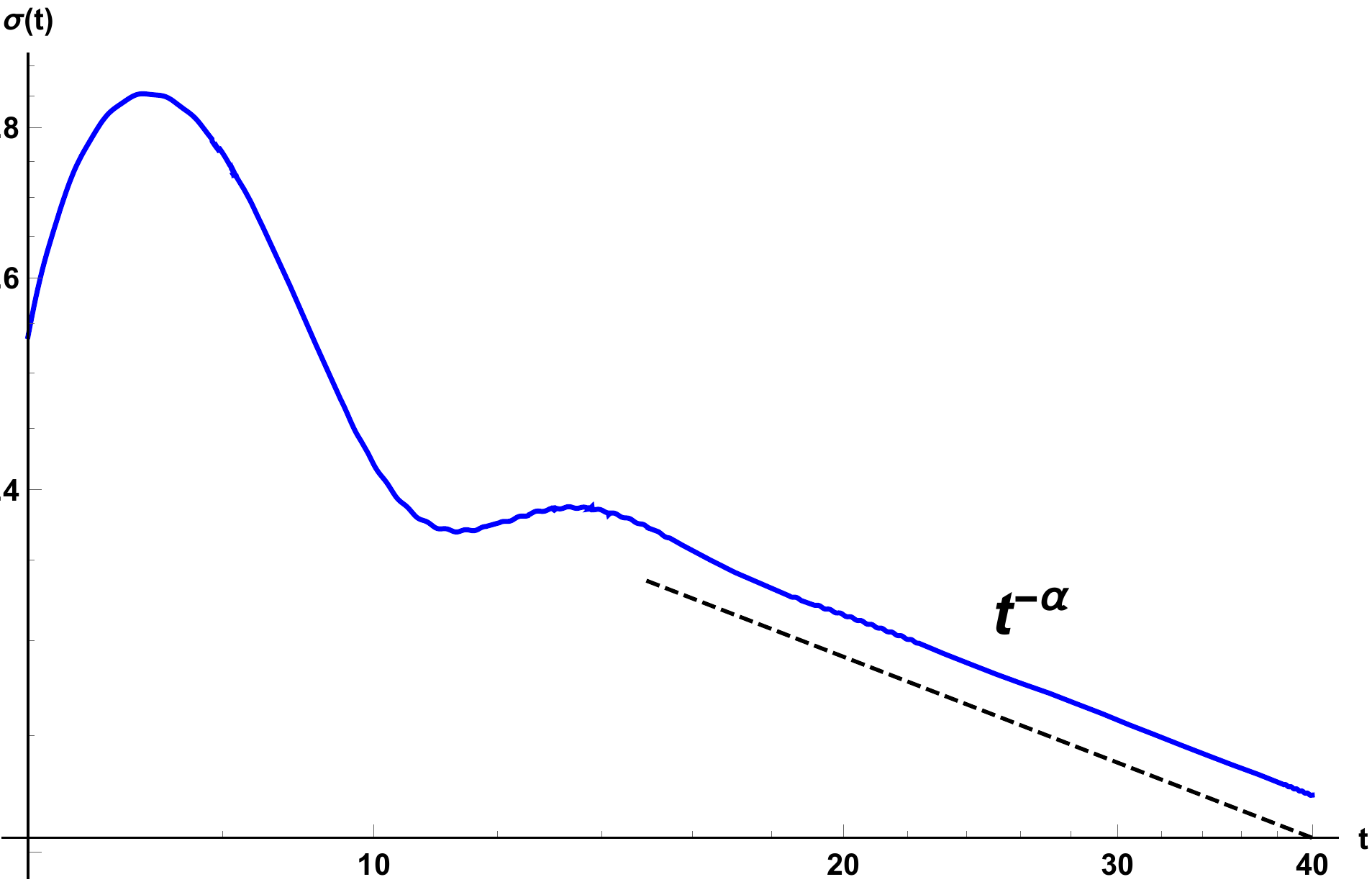}
\caption{The response to a Logistic function strain with $\beta=5$. We consider a single QNM mode along with a branch 
cut. We fix the parameters to $c=9.412,\Gamma=1,m^2=1,a=1,\alpha=0.5,b=5$. \textbf{Left: }The real time response. \textbf{Right: }A log-log plot indicating the late time $t^{-\alpha}$ power law decay.}
\label{figtoy2}
\end{figure}

We will specialize to the fluid case by choosing the constant $c$ that controls the shear elastic modulus:
\begin{equation}
\mu\,=\,Re\left[\mathcal{G}_{toy}(0)\right]
\end{equation}
\noindent such that $\mu=0$. The power $\omega^\alpha$ produces a branch cut on the imaginary axis starting from the 
origin. 
An example of the pole structure is shown in fig.\ref{figtoy1}.
To illustrate the capabilities of this toy model, we consider the response to a step function at strictly zero temperature. 
We display our results for a specific choice of parameters in fig.\ref{figtoy2}. 
We see that our model captures well the essential aspects of the full holographic model shown in fig.\ref{fig:real_time_Z2}, 
showing that the response is dominated by the first QNM. \\

Moreover, the presence of the branch cut $\sim \omega^\alpha$ accounts for the expected power law decay at $T=0$. 
In fact, considering only the branch cut term, which dominates at low frequencies, 
and the fact that the step function is constant at late times, we can derive the scaling:
\begin{equation}
\sigma(t)\,\sim\,\int\,e^{-\,i\,\omega\,t}\,w^{\alpha\,-\,1}\,d\omega\,\sim\,t^{-\,\alpha}
\end{equation}
at late times. This is indeed consistent with the numerical results shown in fig.\ref{figtoy2}. 
Moreover we see that at $T=0$ the late time dynamics is controlled by the branch cut formed 
on the imaginary axis, and not by the gapped QNM, which would yield exponential decay.\\

The toy model is also useful to describe the stress overshoot phenomenon present in the start-up 
experiment. Indeed, truncating to a single resonance (with non-zero real part),
\begin{equation}
\mathcal{G}_{toy2}(\omega)\,=\,c\,+\,\frac{b}{\omega^2\,+\,i\,\omega\,\Gamma\,-\,m^2} \label{toy2}
\end{equation}
we can fix the constant $c$ so as to mimic a fluid model where $\mu=G(0)=0$. This fixes $c=b/m^2$. Note that the parameter $\Gamma$ in this model is the imaginary part of the QNM location. 

In this model it is very easy to obtain the elastic modulus and  viscosity,
\begin{equation}
G_{\infty}\,=\,\frac{b}{m^2},\quad \eta\,=\,\frac{b\,\Gamma}{m^4}
\end{equation}
meaning that the Maxwell relaxation time in this case is:
\begin{equation}
\tau_M\,\equiv\,\frac{\eta}{G_{\infty}}\,=\,\frac{\Gamma}{m^2}
\end{equation}
This now can be compared to the relaxation time encoded in the QNM, 
$$
\tau_{QNM}=1/\Gamma~.
$$ 
Therefore in this simple model, the ratio $\tau_M / \tau_{QNM}  = (\Gamma / m)^2 $, which can be chosen to be large or small.

\begin{figure}
\centering
\includegraphics[width=0.8 \linewidth]{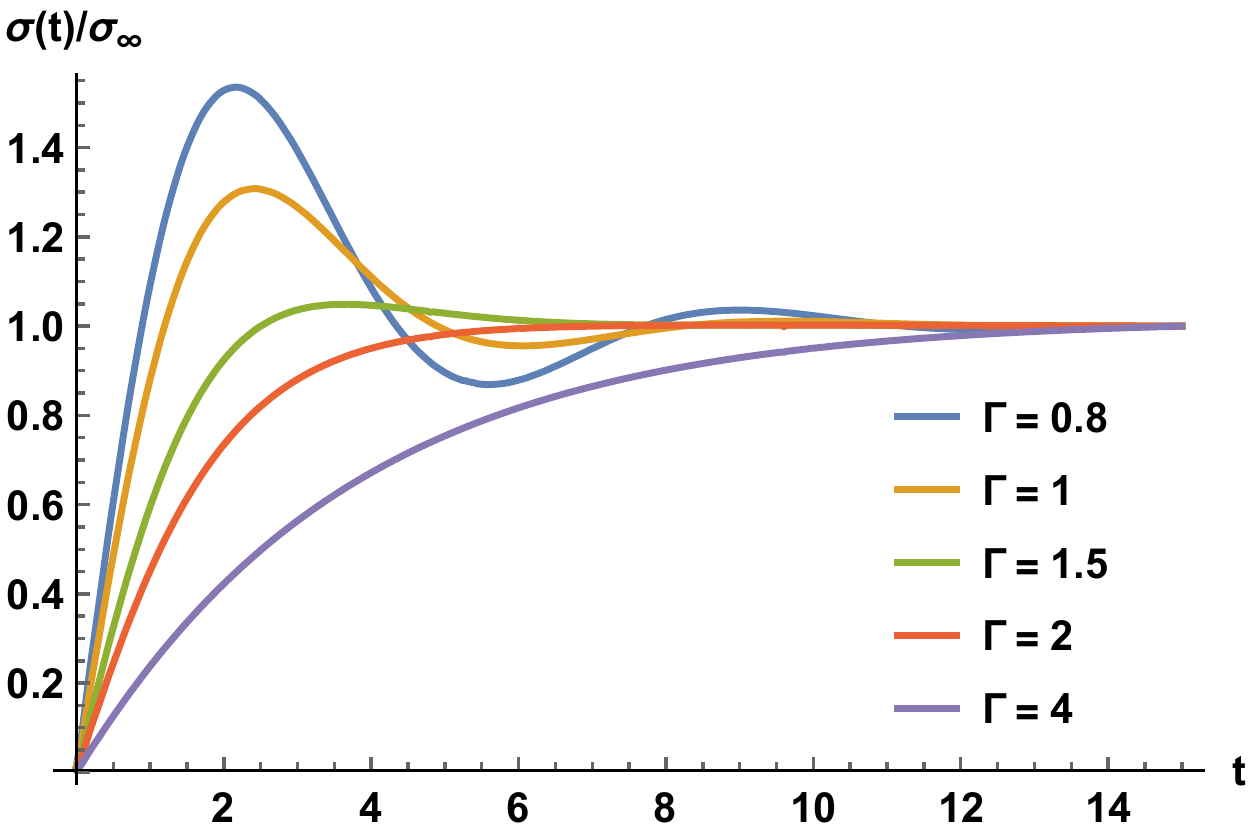}
\caption{The response of the system described by the toy model Green function \eqref{toy2} to a ramp source. We fix $m=1$ and we move $\Gamma$.}
\label{ramptoy}
\end{figure}

Now we can simply compute the response of the system to a ramp strain. The results are shown in fig.\ref{ramptoy}.
Clearly, the overshoot is present for $\Gamma \ll m$, that is 
$$
\tau_M < \tau_{QNM}~.
$$
This is the same pattern that is seen in the fluid models of Sec.~\ref{sec:Z}, and as a matter of fact it will also be found in the solid models below (concerning the overshoot phenonenon):  
materials that develop a {\it light} QNM exhibit the overshoot phenomenon. Here by `light' we mean that the damping scale of the 
QNM is smaller than the relaxation scale encoded in the material moduli (elasticity, shear). 

A rather simple picture seems to emerge, then: in presence of  such a light QNM, the approach to steady state is delayed and moreover it can go together with oscillatory features in the response expected from the fact that one is exciting a really dynamical mode in the system. This picture seems conceptually quite different from other models or interpretations that ascribe the overshoot phenomenon to a superposition of elastic and viscous behaviours. It is clear in our models that the feature behind overshoot is a light QNM.  
It is of course interesting to understand under what conditions does a QNM go below the Maxwell relaxation time. In our solutions, this seems granted because by decreasing temperature the poles on the imaginary axis must get eventually arbitrarily light. In turn, this is a consequence of the fact that our holographic solutions always display criticality at $T=0$ (an AdS${}_2$ extremal horizon).



%

\subsection{Viscoelastic solids}\label{sec:X3}

\begin{figure}
\centering
\includegraphics[width=0.8\linewidth]{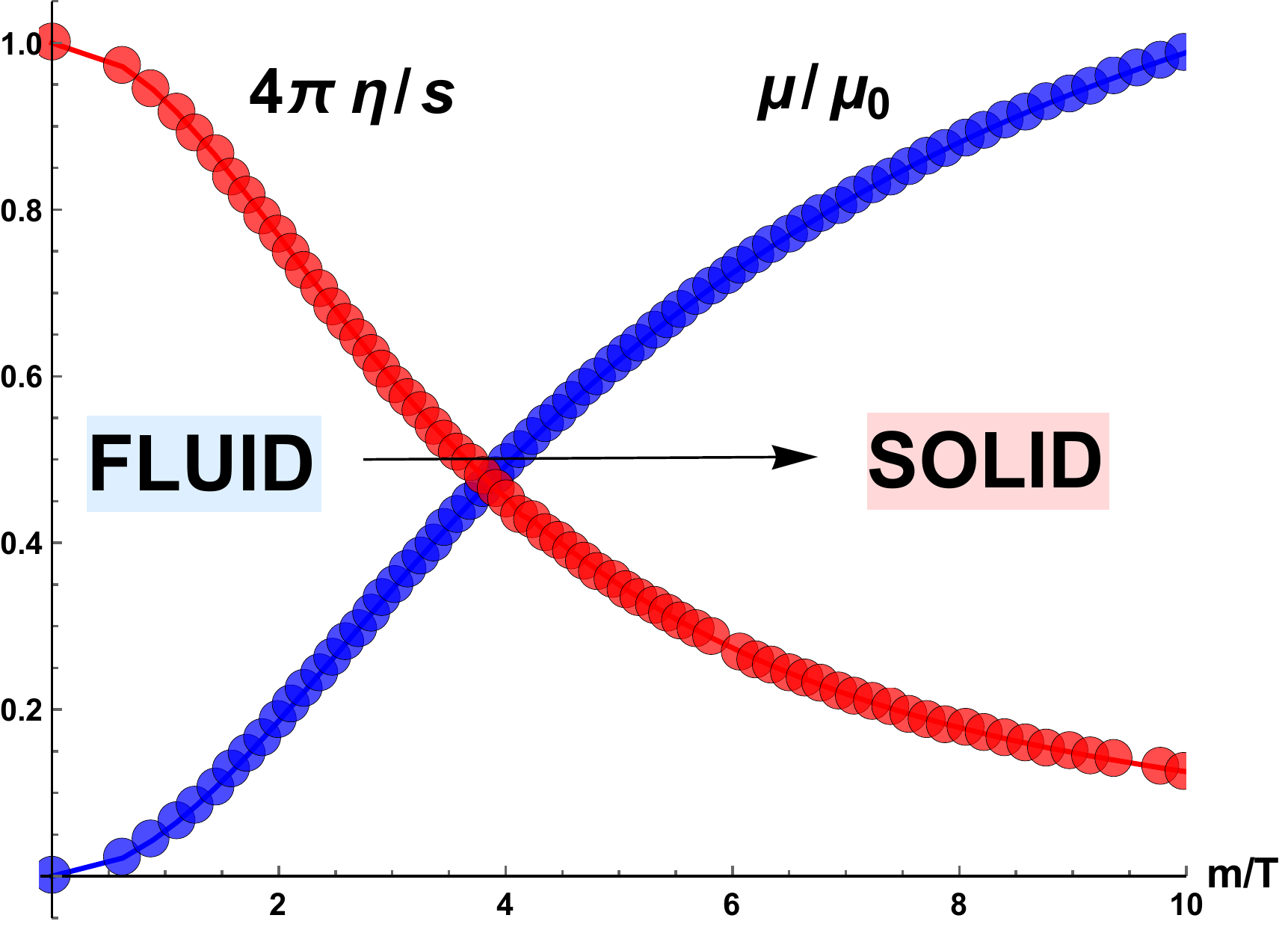}
\caption{A schematic phase diagram of our holographic models taken from the data of the potential $V(X)=X^3$. At $m/T \rightarrow \infty $ the shear modulus is maximal $\mu\equiv \mu_0$ and the viscosity $\eta$ is zero; at $m/T=0$ the viscosity is maximal $\eta/s=1/4\pi$ and the shear modulus is zero. In the intermediate range the dual field theory is viscoelastic with both finite shear modulus and viscosity. The transition appears smooth reminding of the glassy transition in amorphous solids and glasses.}
\label{phasefig}
\end{figure}


In this section we consider a different class of model, which we denote as viscoelastic solid because of the 
presence of a finite static shear modulus $\mu \neq 0$.
More specifically, we study potentials of the form $V(X)=X^n$, which realize the spontaneous symmetry breaking of translational invariance 
and exhibits viscoelastic features along with the presence of damped phonons \cite{Alberte:2017cch,Alberte:2017oqx}. 
As previously analyzed in \cite{Alberte:2015isw,Alberte:2016xja,Alberte:2017oqx}, the shear elastic modulus 
and the shear viscosity coefficient of a class of holographic models of the form \eqref{action},
exhibit a viscoelastic behaviour and a smooth glassy transition. More precisely, the shear modulus $\mu$ 
goes smoothly to zero increasing the dimensionless parameter $T/m$ and it reaches its maximum value 
at zero temperature. The viscosity behaves exactly in the opposite way. For a schematic representation see 
fig.\ref{phasefig}. The holographic systems behave as a perfect dissipationless solid at $T=0$, as a strongly 
coupled viscous fluid at $T \rightarrow \infty$ and as viscoelastic materials in the intermediate range. As a 
consequence, dialling the dimensionless parameter $m/T \in [0,\infty]$ we can study our background in 
the viscoelastic regime.
For brevity, concentrate on the case $n=3$. We have checked that higher power-law potentials $V(X)=X^n$ behave 
in a qualitatively similar fashion. 

\begin{figure}[th]
\centering
\includegraphics[width=0.8 \linewidth]{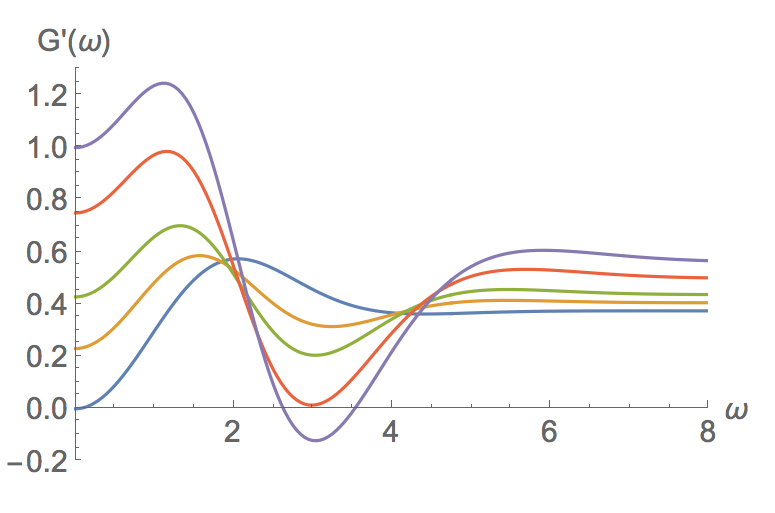}

\vspace{0.4cm}

\includegraphics[width=0.8 \linewidth]{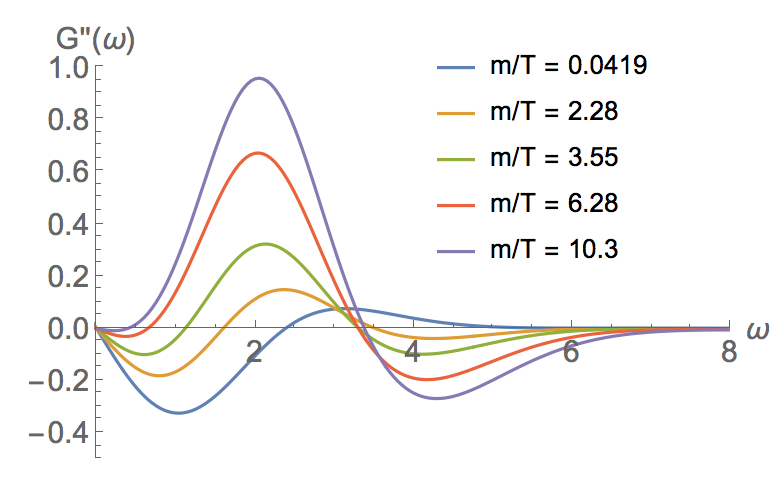}
\caption{Green's function as a function of frequency for $V = X^3$ for 
various values of $m/T$. The shear modulus $\mu=G'(0)$ appears to be non zero at finite $m/T$.}%
\label{fig:G_X3}
\end{figure}

We start by studying the dynamical modulus $G$ in the frequency domain. We show our results
for the frequency dependent Green's function in fig.\ref{fig:G_X3}. 

First, note that our numerics reproduce the generic expansion at low frequencies of the form \eqref{hydro} so that one can readily identify the shear viscosity and elasticity moduli. Note that the main difference  with respect to the previous fluid case is that now
$$
\mu = G(0) \neq0~,
$$
signalling that the model now behaves as a solid in the conventional sense (from the static response).
It can be appreciated in fig. \ref{fig:G_X3} that this modulus drops quite quickly to zero at high temperatures, corresponding to a melting-like transition (see \cite{Alberte:2017oqx,Alberte:2017cch}). In fact the whole functional dependence of $G'$ and $G''$ in the high $T$ limit resemble the one in the fluid limit, fig. \ref{fig:G_Z2}.

Secondly, at large frequencies $\omega \rightarrow \infty$ the imaginary and dissipative 
part of the complex moduli goes to zero while the real and elastic part asymptotes a constant value, $G_\infty$, 
again given by the energy density as in \eqref{Ginf E}.
This immediately means that there are (at least) two notions now of elastic modulus, and therefore of 
 {\it Maxwell relaxation times}:
\be\label{tMs}
\tau_M^{(\infty)} = \eta / G_\infty \quad\quad {\rm and} \quad\quad  \tau_M^{(0)} = \eta / \mu~.
\ee
On general grounds, it is natural to expect that the relevant one at high temperatures is  $\tau^{(\infty)}_M$, as in Sect.~\ref{sec:Z}, because in this limit material melts, meaning that $\mu$ must vanish and the fluid behaviour must be recovered. Conversely, at low temperatures the a static elastic modulus $\mu$ is expected to be sizeable. Then, the late time response is encoded directly in the low frequency expansion of the Green's function \eqref{hydro}, suggesting that the relevant relaxation time is should be given by $\tau^{(0)}_M$ (at low $T$). In any case, a glance at the left panel of fig. \ref{fig:G_X3} reveals that in our models at low temperatures $\mu$ and $G_\infty$ are comparable in magnitude so there isn't a big difference in the two Maxwell times in this regime.

\begin{figure}[th]
\centering
\includegraphics[width=0.8 \linewidth]{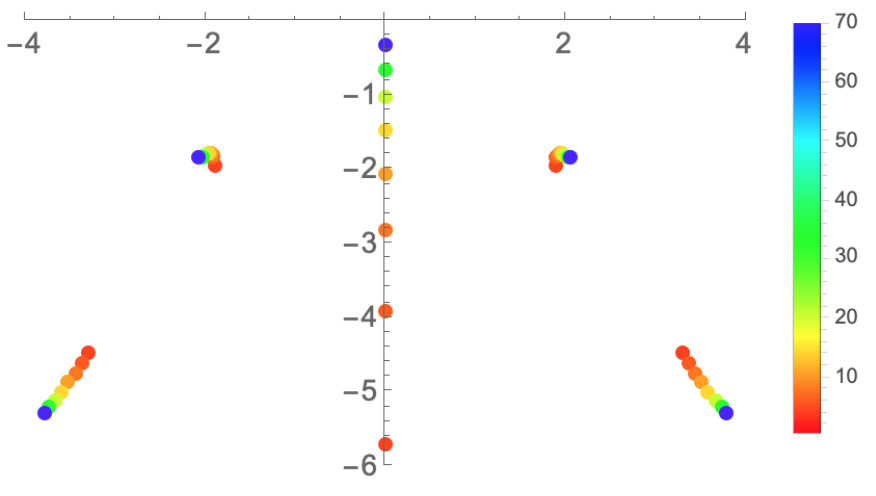}

\vspace{0.4cm}

\includegraphics[width=0.8 \linewidth]{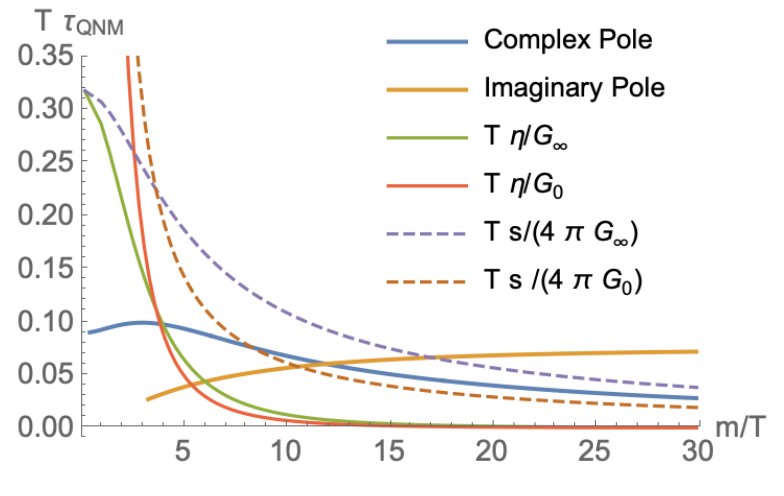}
\caption{{\bf Top:} QNMs in the complex plane for $V=X^3$ and varying $m/T$. To ease visualization, 
we only show the purely imaginary QNMs which is closest to the real axis. 
{\bf Bottom:} 
Relaxation associated to the lightest complex and purely imaginary QNMs and relaxation times Maxwell relaxation times \eqref{tMs}, as a function of $m/T$. 
At  $m/T\gtrsim4$, both Maxwell relaxation times become (remarkably simultaneously) smaller than the QNM relaxation time scale (that is set by the QNMs). The dotted line in the Right panel shows that the QNM relaxation time follows quite closely $s/(4\pi G_{0,\infty})$ rather than $\eta/G_{0,\infty}$.}%
\label{fig:QNM_X3}
\end{figure}

Next, we observe that the position of the dominant finite frequency peak 
in $G'$ is approximately given by the real part of the leading gapped QNM, see 
fig.\ref{fig:QNM_X3}.
Like in the fluid model, there are two `light' QNMs: a pure imaginary one and an off-axis mode, and the at sufficiently low $T$ the imaginary mode is granted to have lowest imaginary part because $T = 0$ the accumulation of poles in the imaginary axis signals the formation of a branch cut. In fact, one can estimate the temperature at which  the cross-over between the dominance of the gapped QNM and the branch cut by comparing the imaginary parts of 
the first propagating and damped QNMs, see right panel of fig. \ref{fig:QNM_X3}.

\begin{figure}[th]
\centering
\includegraphics[width=0.8 \linewidth]{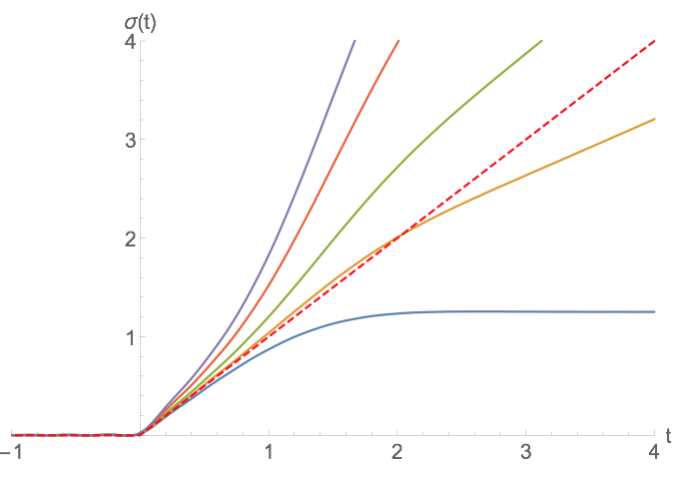}

\vspace{0.4cm}

\includegraphics[width=1 \linewidth]{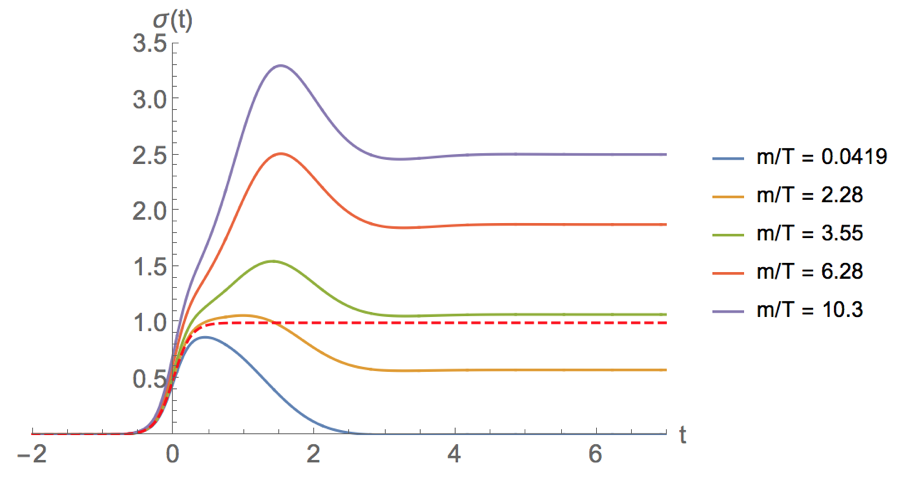}
\caption{Real time responses for the potential $V=X^3$, for a ramp source (top) and a logistic source (bottom). Just like for the fluids the overshoot behaviour is found at $m/T\gtrsim 3.5$, which coincides with  Maxwell being smaller than QNM relaxation time.}%
\label{fig:real_time_X3}
\end{figure}

It is relevant to compare the relaxation times  derived from the lowest of these QNM modes,
$\tau_{QNM}\,=\,-\, 1/ Im\left(\omega_{QNM}^*\right)$, to the above Maxwell relaxation times $\tau_M^{(0)}$, $\tau_M^{(\infty)}$.
As we can see in the right panel of fig.~\ref{fig:QNM_X3}, both $\tau_M^{(0)}$ and $\tau_M^{(\infty)}$ go below the $\tau_{QNM}$, towards low temperature. In fact, they do so at a  surprisingly coincident temperature, roughly $m/T\simeq 4$.  
Since this crossover takes place at smaller $m/T$ as compared to the fluid case of Section \ref{sec:Z}, the QNM responsible for this crossover is the  complex one in the $V(X)$ model. At higher $m/T$, there is a second crossover where the imaginary QNM becomes longer lived than the complex one. However, the first crossover at $m/T\simeq4$ is also seen to correlate well with the appearance of overshoot in the step-function external source, as is especially clear in the right panel of  fig.~\ref{fig:real_time_X3}.

Another significant difference with respect to the fluid case is that the temperature dependence of the relaxation times from the QNMs $\tau_{QNM}$ or from the Maxwell formulas \eqref{tMs} differ substantially: now both $\tau_M^{(0,\infty)}$ decrease much faster towards $T=0$ than $\tau_{QNM}$. 
It is natural to associate this to the fact that the solid models that we are considering now are known to violate \cite{Alberte:2016xja,Hartnoll:2016tri} the KSS bound \cite{Kovtun:2004de}. In these models $\eta/s$ decreases `abnormally' at low $T$ (while $s$ remains nonzero) so perhaps this is the only reason why the Maxwell time does too. More specifically, since viscosity and entropy density have the same units, it is also possible to construct a relaxation time out of the ratio 
$$
\tau_s=\frac{s}{4\pi\,G}
$$
with $G$ denoting any of $G_\infty$ or $G(0)$ and the $4\pi$ factor is introduced in order to match correctly the fluid limit (high temperature). As shown in the right panel of fig.~\ref{fig:QNM_X3} the temperature behaviour of $\tau_s$ tracks quite well the relaxation times from the lowest QNM. We leave a further study of this for later investigation.

%

We display our results for the real time response in fig. \ref{fig:real_time_X3}. 
Let us discuss first the ramp source. In this case, at small $m/T$ the system reacts like a viscous fluids where the strain $\sigma(t) \sim \dot{\epsilon}(t) = const.$; nevertheless before reaching the constant value there is a time delay of the system with a 
characteristic timescale $\tau_d$.
Once $m/T$ is increased, a finite shear modulus $\mu \neq 0$ appears and therefore the response 
continues to grow linearly at late times.
Notice that the slope of the late time response indeed grows with $m/T$ as the shear modulus does. 
The dissipative response provides only a small deviation from the linear growth which is visible in a transient regime. 
In realistic materials once the strain becomes large, \textit{i.e.} at late time, non-linear effects become important. 
As a consequence what usually happens is that the rigid bonds in the material become weaker and the stress stops to grow 
and it reaches the so-called \textit{Newtonian plateau} which for example we see at $m/T=0$. Clearly these effects cannot be captured by our analysis which is limited to the linear response approximation. 
In some cases non-linear effects produce overshooting \cite{PhysRevB.90.140203}. It  would be interesting to see if this effect
 can be captured by a holographic calculation. We hope to come back to this interesting point in the future.  

\begin{figure}[th]
\centering
\includegraphics[width=0.8 \linewidth]{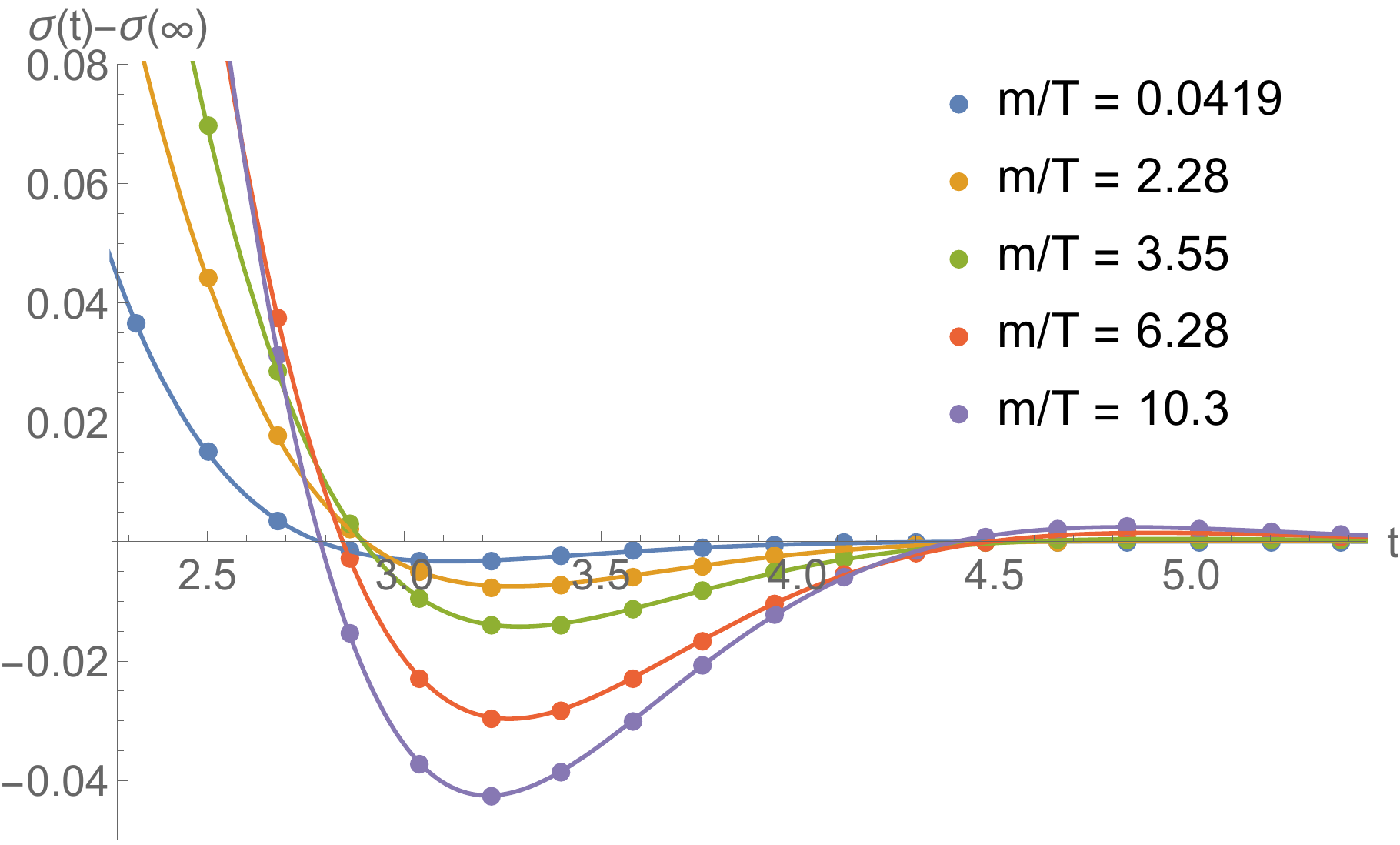}
\caption{Fit of the late time dynamics in the case of the logistic response for the $V = X^3$ 
model to the lowest QNMs. The data points mark our numerical data, while the solid lines correspond 
to the QNM fit. To ease visualization we have subtracted the equilibrium value.}%
\label{fig:late_X3}
\end{figure}

We now turn to the step function. In that case at $m/T=0$ the response is purely viscous: it develops a peak when the gradient of the strain is large (at the step position) and then it dies off as expected since at late time $\dot{\epsilon}(t)=0$. On the contrary, switching on $m/T$ the shear modulus is non-vanishing and so that the response at late time asymptotes a constant proportional to the shear modulus. 
Before reaching such a plateau a peak appears in the response due to the competition between the viscous and elastic response.
The peak is sharper and higher the stronger is the elastic response, which is governed by the  resonance in the QNMs spectrum. 
The residue of such resonance controls the strength of the peak and its imaginary value the exponential decay that follows 
it towards the late time plateau. 
In fact, we have checked that the approach to equilibrium is governed by the first QNM, see fig \ref{fig:late_X3}. 
In the range of temperatures we have considered, we obtain a good fit considering the lightest QNM only. 

\section{Conclusions}\label{sec: conclusions}

We have studied the real time mechanical response of a class of solid and fluid holographic massive gravity models. 
For simplicity we have restricted ourselves to the shear sector and  to consider the linear response approximation valid 
only for small external sources.
Within this framework we have obtained the complex moduli $G(\omega)$ from the Green's 
function of the stress tensor operator $T_{xy}$ at finite frequency. Our results support the classification of 
the models at hand into a solid and a fluid class accordingly to the value of the shear modulus $\mu \equiv G(0)$. 


We have computed, within linear response, the real time response of the systems under an applied external time dependent strain. The results show interesting features which are typically observed in viscoelastic materials:  
stress overshoot and stress relaxation.
%


%

Our results have a number of implications:
\begin{enumerate}
\item They confirm that the holographic massive gravity models \cite{Baggioli:2014roa,Alberte:2015isw} as dual bulk descriptions of (strongly coupled) viscoelastic materials, either viscoelastic solids or fluids depending on the choice of potential.

\item Just like in the phenomenological Maxwell, model these models possess a well defined elastic modulus defined as   $G_\infty = \lim_{\omega\to\infty}G(\omega)$  that is a key player in the viscoelastic phenomena and which is nonzero even  fluids (which have $G(0)=0$). Moreover, in our models we find that this is basically given by the energy density, see Eq.~\eqref{Ginf E}.

\item At finite temperature, the relaxation time is parametrically similar to the Maxwell relaxation time $\tau_M=\eta/G_{\infty}$, but in close inspection there are important differences between this and the actual relaxation time. At high temperatures $\tau_M$ is only a factor $\sim 3$ different (larger) and moreover it captures correctly the temperature dependence. However, at lower temperatures, the $T$ behaviour starts to differ and at small enough $T$,  $\tau_M$ severely underestimates $\tau_{rel}$. In our models, this must happen because dynamics at  $T=0$ is controlled by an infrared critical point, and the relaxation in time must be power-law. 
It is tempting to speculate whether a similar behaviour holds more generally -- that whenever the mechanical deformation of a material is approaching a critical regime, then the Maxwell relaxation time $\tau_M$ must necessarily underestimate the actual relaxation time. 

\item We find that the appearance of a stress overshoot peak preceding relaxation correlates well with the underestimation condition $\tau_M < \tau_{rel}$. As we have argued, this should be granted to occur in  materials that are close to criticality. It would be interesting to understand whether this can occur in non-critical materials, via other mechanisms  \cite{divoux2011stress,PhysRevLett.118.018002,PhysRevB.90.140203}, or whether they are two sides of the same reality.

%
%


\end{enumerate}

One interesting and natural future direction is extending our computations to the non-linear regime. The latter could produce 
more realistic features at late time and shed more light on the nature of the stress overshoot in strongly coupled viscoelastic materials. 
One possibility is to follow the methods used for the non-linear eletric conductivity in \cite{Withers:2016lft,Baggioli:2016oju} or
 take inspiration from the models for non-linear elasticity developed in \cite{Alberte:2018doe,coming}.

It is interesting to compare our power law relaxation with the existing literature. A viscoelastic model that predicts a power-law scaling via a fractional relaxation process has been introduced in \cite{pritchard2017oscillations} and analyzed further in \cite{PhysRevE.95.023001}. 
There is some evidence that there are viscoelastic materials, like polymeric media, biological tissues and cells 
(see \cite{METZLER2003941,Shen2013} for some examples), where the power law scaling is at work. 
%
See  \cite{LaNave:2017lwf,Limtragool:2016ghz,vlahos2008normal} for an incomplete list of references that use AdS/CFT techniques to model phenomena where power-law scaling is important.\\ 
 
As a final and promising direction we have to mention the recent interest around the study of phonons and glassy/viscoelastic features in the context of quantum critical scale invariant systems. In particular two important questions arised: what is the behaviour of the phonons at quantum criticality \cite{ishii2019glass}? How do viscoelastic properties affect the onset of high-T$_c$ superconductivity \cite{2019arXiv190200516S}? Preliminary results suggest a strong correlation between phonons at criticality, viscoelasticity and glass-enhanced high-T$_c$ superconductors. The methods presented here seem useful to address some of these questions. We hope to revisit these issues in the future.


\section*{Acknowledgements}

We thank Kostya Trachenko and especially Alessio Zaccone for useful discussions and comments about this work and the topics considered. We would like to thank Alessio Zaccone for explaining us all the secrets of viscoelastic materials and the related open problems at any time of the day and the night.
The work of TA is supported by the ERC Advanced Grant GravBHs-692951. 
MB acknowledges the support of the Spanish MINECO’s “Centro de Excelencia Severo Ochoa” Programme under grant SEV-2012-0249. 
OP acknowledges support by the Spanish Ministry MEC under grant FPA2014-55613-P, FPA2017-88915-P the Severo Ochoa excellence program of MINECO (grant SO-2012-0234, SEV-2016-0588), as well as by the Generalitat de Catalunya under grants 2014-SGR-1450, 2017SGR1069.
TA gratefully acknowledges the hospitality of the Lorentz Center during the 
workshop ``Bringing Holography to the Lab: Explaining Strange Metals with Virtual Black Holes", 
held in Leiden, 2019, where part of this work was carried out. 
MB would like to thank Marianna Siouti for the unconditional support.
MB would like to thank IFAE and UPC Barcelona for the warm hospitality during the completion of this work. We would also like to acknowledge IFAE for providing the comfortable atmosphere where this project started.

\appendix

\section{The instantaneous elastic modulus and the energy density}
\label{app:ginf}

In this appendix we obtain the large frequency result for the two point function of the stress tensor. 
Let us start with the generic equation for the shear mode:
\begin{equation}
h_{xy} \left(-\frac{2 m^2 V_X}{f}-\frac{2 i \omega}{u f}\right)+h_{xy}' \left(\frac{f'}{f}+\frac{2 i \omega}{f}-\frac{2}{u}\right)+h_{xy}''=0
\end{equation}
and let us write down the blackening factor in the general form in terms of the energy density $\epsilon$ (setting the radius of the horizon to $u_h=1$) :
\begin{equation}
f(u)\,=\,1\,-\,\epsilon\,u^3\,+(\epsilon\,-\,1)\,u^n
\end{equation}
which is always the case for the models we consider. The specific value of $n$ is actually not relevant for the following derivation but importantly $n \neq 3$. If we for example choose a potential of the type $V(X,Z)=X^{p_1}Z^{p_2}$ then we have $n=4 p_2+2p_1$.\\
Imposing regularity at the horizon, we find that the series solution near $u=1$ has the form
\begin{equation}
 \label{hIR}
	h_{xy} = h^H - h^H (1- u) + O(1/\omega) 
\end{equation}
On the other hand, the general near boundary expansion has the form
\begin{equation}
\label{hUV}
	h_{xy} = h^{(0)}(1  + i \omega  u  + h^{(3)} u^3 + \omega h^{(4)} u^4 + \omega^2 h^{(5)} u^5 + h^{(6)} u^6  + \ldots   )
\end{equation}
\noindent where
\begin{equation}
	h^{(4)} = c_0 ( h^{(3)} - A h^{(0)} ) , \qquad h^{(5)} = c_1 ( h^{(3)} - A h^{(0)} ) 
\end{equation}
\noindent where we have made the $\omega$ dependence explicit, and $c_0$, $c_1$ some numerical $\omega$-independent
constants, and 
\begin{equation}
	A = - \frac{1}{4} \,\epsilon
\end{equation}
From \eqref{hIR}, we observe that an appropriate approximate solution in the limit $\omega \to \infty$ has the form
\begin{equation}
\label{h approx}
	h = u + \frac{i}{\omega} \Delta (u) 
\end{equation}
Here, we have set $h^H = 1$ which is always valid by linearity of the equation. In order for \eqref{h approx}
to hold approximately everywhere, we choose 
\begin{equation}
	h^{(0)} = \frac{i}{\omega}, \qquad h^{(3)} = A h^{(0)}
\end{equation}
Note that there could be subleading contributions in powers of $u$ in \eqref{hUV}, however, these would be
suppressed since $0<u<1$. We have verified numerically that indeed, for large $\omega$ our solutions have 
the form \eqref{h approx}. We can read off the Green's function from this approximate solution, which yields, 
\begin{equation}
	G_\infty = - \frac{3}{2} A =  \frac{3}{8} \epsilon
\end{equation}
\noindent as promised.\\
We suspect this is a universal relation for holographic systems which we plan to investigate in the nearest future. 
\section{IR conformal dimension of the shear operator}\label{appIR}
In this section we compute the conformal dimension of the shear operator $T_{xy}$ at the IR fixed point or in other words at the extremal near horizon geometry AdS$_2 \times R^2$.\\
As shown in the main text the UV conformal dimension of the stress tensor does not get modified by the graviton mass. The radial dependent graviton mass indeed can be defined as:
\begin{equation}
m_g^2(u)\,\sim\,m^2\,u^2\,V_X
\end{equation}
which indeed is always zero at the UV boundary $u=0$. As a consequence the conformal dimension of the stress tensor is the canonical one, in this case $\big[T_{xy}\big]=3$.\\
At the zero temperature IR fixed point the situation is different. Let us consider the extremal near-horizon geometry:
\begin{align}
ds^2\,=\,\frac{L^2}{u_0^2}\,\Big[&dx^2\,+\,dy^2\,-\,\frac{f''(u_0)}{2}(u-u_0)^2\,dt^2\,\nonumber\\
&+\,\frac{2}{f''(u_0)\,(u-u_0)^2}\,du^2\Big]
\end{align}
where $u_0$ is the position of the extremal horizon and $L$ the radius of the AdS$_4$ manifold. It is simple to realize that the above geometry is an AdS$_2 \times R^2$ spacetime where the radius of the AdS$_2$ geometry reads:
\begin{equation}
L_2^2\,=\,\frac{2\,L^2}{f''(u_0)\,u_0^2}
\end{equation}
For the models we consider we find:
\begin{equation}
f''(u_0)\,=\,\frac{2\,m^2\,L^2}{u_0^2}\,\left(X\,V_X\,+\,2\,Z\,V_Z\right)\Big|_{u=u_0}
\end{equation}
Now we can solve the equation of motion close to the horizon obtaining:
\begin{equation}
h_{xy}\,=\,\alpha\,(u-u_0)^{p_{+}}\,+\,\beta\,(u-u_0)^{p_{-}}
\end{equation}
where:
\begin{equation}
p_\pm\,=\,\frac{1}{2}\,\left(-\,1\,\pm\,\sqrt{4\,M^2\,+\,1}\right)
\end{equation}
and:
\begin{equation}
M^2\,\equiv\,\frac{4\,m^2\,V_X}{f''(u_0)}
\end{equation}
Putting all the results together and assuming a potential of the form $V(X,Z)=X^a Z^{(b-a)/2}$ we obtain\footnote{The consistency of the model (no instabilities, positive elastic moduli, etc) imposes that the parameters satify $a>0$, $b>3/2$ \cite{Alberte:2018doe,coming} }:
\begin{equation}
M^2\,=\,\frac{2\,a}{b}
\end{equation}
The conformal dimension $\Delta$ of the $T_{xy}$ operator is then simply:
\begin{equation}
\Delta\,=\,1\,+\,p_{+}\,=\,\frac{1}{2}\,+\,\frac{1}{2}\,\sqrt{\frac{8\,a}{b}\,+\,1}\label{dimdim}
\end{equation}
In the case of the $X$ dependent potential $a=b$ we obtain $\Delta=2$ which is in agreement with the result for the linear potential presented in \cite{Hartnoll:2016tri}.\\
The just derived conformal dimension is governing two different phenomena:
\begin{itemize}
\item The late time power-law relaxation of the stress $\sigma(t)$ which appears at $T=0$. Hereby we sketch the simple argument behind it. The Green function for an operator with dimension $\Delta$ does scale like:
\begin{equation}
G_{\Delta}(\omega)\,\sim\,\omega^{2\Delta\,-\,1\,-\,D/z}
\end{equation}
where $D$ is the number of spatial dimensions and $z$ the Lifshitz scaling parameter.
In our case $D=2$ and $z$ is formally infinite because of the AdS$_2$ factor in the near-horizon extremal geometry. In few words the Green function of the stress tensor operator, which controls the linear stress-strain response, scales like $\sim \omega^{2 \Delta -1}$ with the conformal dimension obtained above in \eqref{dimdim}. By dimensional analysis we can then simply deduce that at late time the stress will relax following the power law scaling:
\begin{equation}
\sigma(t)\,\sim\,t^{-\sqrt{1\,+\,\frac{8\,a}{b}}}
\end{equation}
which is always faster than the linear scaling $t^{-1}$ which appears at zero graviton mass $V_X=0$ or equivalently $a=0$. Notice that this implies:
\begin{align}
&\sigma(t)\,\sim\,t^{-1}\,\quad \text{for} \quad V(X,Z)=Z^q\\
&\sigma(t)\,\sim\,t^{-3}\,\quad \text{for}\quad V(X,Z)=X^N
\end{align}
at zero temperature.
\item The power-law fall-off of the $\eta/s$ ratio at zero temperature:
\begin{equation}\label{etas}
\frac{\eta}{s}\,\sim\,T^{2\,p_{+}}\,=\,T^{-1\,+\,\sqrt{\frac{8\,a}{b}\,+\,1}}~.
\end{equation}
Notice that indeed for $V_X=0$ ($a=0$) this gives that the $\eta/s$ ratio is constant in temperature as it must be. 
Notice also that \eqref{etas} extends the results of \cite{Hartnoll:2016tri}, which found that $\eta/s$ approaches $0$ no faster than $T^2$. Indeed, a power larger than 2 can be obtained for $a>b$, which is compatible with consistency of the model \cite{Alberte:2018doe,coming}. Hence, the fall-off of the viscosity/entropy ratio can be arbitrarily fast without appealing to hyperscaling violation or Lifshitz scaling as in \cite{Ling:2016yxy}.
\end{itemize}

\section{Details about the numerical methods}
In this appendix we provide more details about the numerical techniques used in the manuscript and the derivation of the main results.
\subsubsection*{The Green function}
The stress tensor correlator $\langle T_{xy}T_{xy} \rangle$ is extracted by considering the bulk equation for the shear perturbation
\begin{equation}
h_{xy}\left(-\frac{2 m^2 V_X}{f}-\frac{2 i \omega}{u f}\right)+h_{xy}' \left(\frac{f'}{f}+\frac{2 i \omega}{f}-\frac{2}{u}\right)+h_{xy}''=0\label{EQ1}
\end{equation}
written in EF coordinates. 
The UV asymptotic behaviour of the $h_{xy}$ field close to the boundary $u=0$ reads:
\begin{equation}
h_{xy}\,=h_{xy\,(l)}(\omega)\,(1\,+\,\dots)\,+\,h_{xy\,(s)}(\omega)\,u^{3}\,(1\,+\,\dots)
\end{equation}
and finally the Green function can be read using the standard holographic dictionary as:
\begin{equation}
 \mathcal{G}^{\textrm{(R)}}_{T_{xy}T_{xy}}(\omega)\,=\,\frac{2\,\Delta-d}{2}\,\frac{h_{xy\,(s)}(\omega)}{h_{xy\,(l)}(\omega)}\,=\,\frac{3}{2}\frac{h_{xy\,(s)} (\omega) }{h_{xy\,(l)} (\omega)}\label{green}
\end{equation}
In order to derive this function numerically we use a double shooting matching procedure described in the following steps\footnote{For more details about this method see the lectures notes in \cite{Baggioli:2019rrs}.}
\begin{enumerate}
\item We fix a specific form of the potential $V(X,Z)$ in eq.\eqref{EQ1}.
\item We solve perturbatively equation \eqref{EQ1} close to the horizon $u_h=1$ using an ansatz:
\begin{equation}
h_{xy}^{IR}(u)\,=\,h_0\,+\,\sum_{n=1}^{N}\,h_n\,(1-u)^n
\end{equation}
where we impose regularity and we fix $N=5$ to have enough accuracy. The free parameters of the IR expansion are $\{h_0,\omega,m^2\}$.
\item We construct numerically the solution to equation \eqref{EQ1} by shooting from the IR with the boundary conditions obtained in the previous point to an intermediate radial position $0<u_m<1$ (we concretely choose $u_m=1/2$).
\item We repeat exactly the same procedure from the UV. In this case the free parameters are $\{h_{xy(l)},h_{xy(s)},\omega,m^2\}$.
\item For a generic choice of parameters the two solutions will be not continuous at the intermediate point $u_m=1/2$ as shown in fig.\ref{what}.
\begin{figure}[h!]
\center
\includegraphics[width=0.8\linewidth]{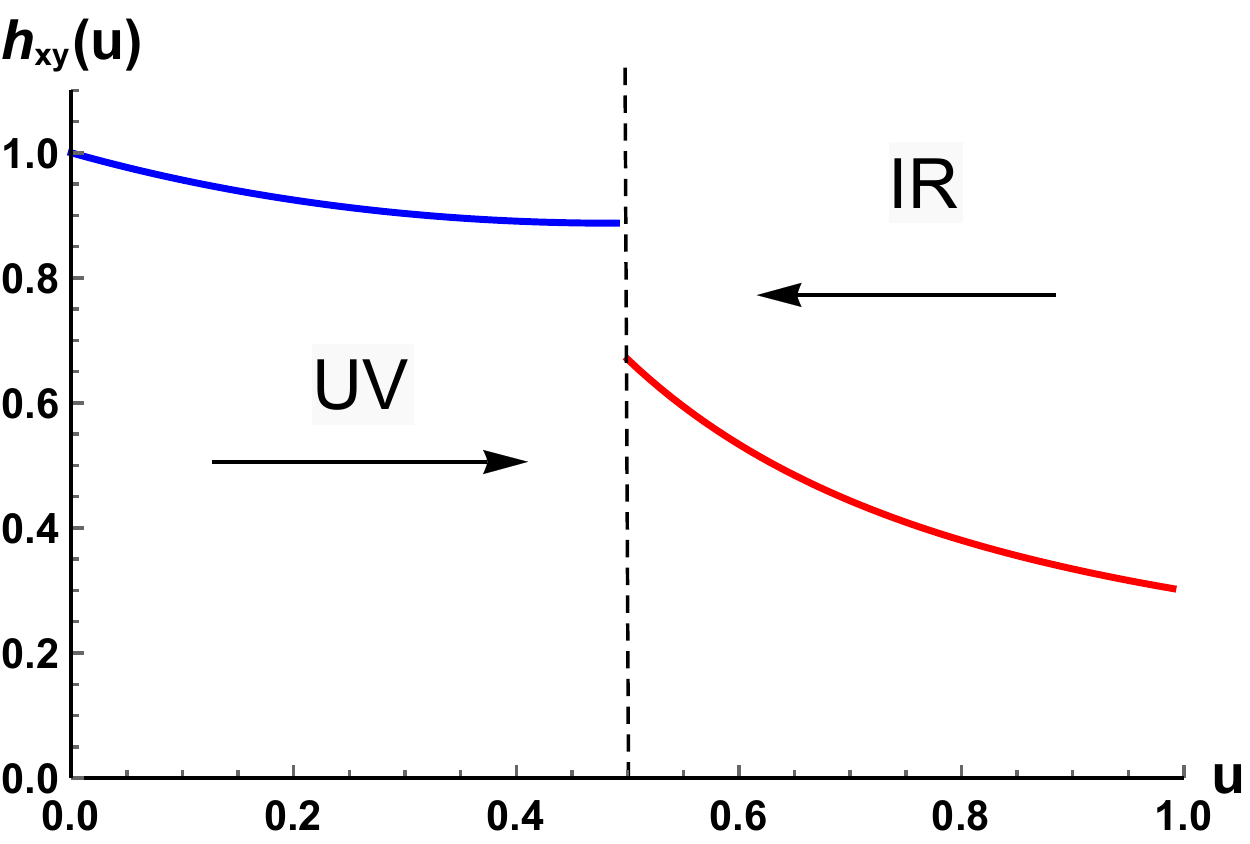}
\caption{The IR and UV solution for a random choice of initial shooting parameters.}
\label{what}
\end{figure}
\item The first step to continue is noticing that one between $h_0$ and $h_{xy(l)}$ is redundant and it can be set to $1$ by rescaling the bulk field $h_{xy}(u)$. After doing that, we have simply to impose the continuity of the function $h_{xy}(u)$ and its first derivative at the intermediate position $u_m=1/2$. This can be easily done with a FindRoot routine in Mathematica. These two requirement will fix for us $h_{xy(l)},h_{xy(s)}$ and therefore our Green's function in terms of $\omega$ and $m^2$.
\end{enumerate}

\subsection*{Quasinormal modes}

The QNMs of the system correspond to the poles of the Green's function. From \eqref{green}, we see that 
this corresponds to setting $h_{xy\,(l)}(\omega) = 0$. Imposing ingoing boundary condition at the horizon, 
\eqref{EQ1} turns into an eigenvalue problem for the frequency $\omega$ which is satisfied by a discrete number 
of complex values for a given set of external parameters. We solve this by discretizing \eqref{EQ1} on a Chebyshev grid
and solving the resulting linear algebra problem numerically. If one wishes, this problem can also be solved by shooting 
as explained in the previous section. 

\subsection*{Late time relaxation and fits}

As it is well-known, the lowest QNMs govern the approach to equilibrium of black holes. In the main text, we have checked that 
this is the case by employing the late time expansion
\begin{equation}
\label{sigma fit}
	\sigma(t) - \sigma(t = \infty) = {\rm Re} \, \left[ \sum_{i=1}^{N_{QNM}} e^{- i \omega_i t} a_i \right]
\end{equation}
\noindent where $N_{QNM}$ is the number of lowest QNMs included in the fit, with frequencies $\omega_i$, the values of
which we have extracted as explained in the previous section. The quantity $\sigma(t = \infty)$ is the constant late time 
value of $\sigma$. 
The complex coefficients $a_i$ are obtained by computing the best fit to the numerical value of the left hand side of \eqref{sigma fit}. 
We have checked the robustness of our results against changing the length of the time interval in which the fits are performed. 
In this simple setting, over-fitting is signalled by large values of some of the coefficients $a_i$. We avoid this by keeping a value of $N_{QNM}$
not higher than 3.

\bibliographystyle{apsrev4-1}

\bibliography{visc}

\end{document}